\documentclass[%
 reprint,
 amsmath,amssymb,
 aps,
floatfix
]{revtex4-2}

\usepackage{graphicx}% Include figure files
\usepackage{dcolumn}% Align table columns on decimal point
\usepackage{bm}% bold math
\usepackage[mathlines]{lineno}% Enable numbering of text and display math
\begin{document}
%\preprint{APS/123-QED}
\title{Impact of feedback time-distribution on laser dynamics}

\author{Martin~Skënderas$^1$}
\author{Spencer~W.~Jolly$^{1,2}$}
\author{Martin~Virte$^1$}
\affiliation{$^1$Brussels Photonics, Vrije Universiteit Brussel, Pleinlaan 2, B-1050 Brussels, Belgium\\
$^2$OPERA-Photonique, Universite Libre de Bruxelles, 50 Av.F.D.Roosevelt, B-1050 Bruxelles, Belgium}

\begin{abstract} 
Time-distributed optical feedback in semiconductor lasers has gained attention for its ability to produce high-quality chaos and effectively suppress the time-delay signature. However, the fundamental impact of the distribution of feedback in time on laser dynamics remains unexplored. In this paper, we investigate this topic by using fiber Bragg grating (FBG) feedback. We theoretically study the laser response using FBGs of different lengths but similar reflectivity, effectively stretching the impulse response over a longer period while maintaining its overall shape. We observe that above a critical value corresponding to a grating length of approximately $1$\,cm, fluctuations in laser stability emerge. We attribute this phenomenon to the damping of relaxation oscillations when the zeros of the FBG reflectivity spectrum align with the laser side lobes around the relaxation oscillation frequency. We also uncover an asymmetrical dynamical behavior of the laser for positive and negative frequency detuning. We deduce that this asymmetry is a characteristic feature of FBG feedback and delve into the specificities that trigger such behavior.  
\end{abstract} 

\maketitle

%\tableofcontents

\section{Introduction} 
\label{sec:intro} 
Semiconductor lasers subject to optical feedback have attracted increasing interest in the past decades, initially as a tool to explore the dynamics of systems with time delay \cite{RevModPhys.85.421}, and later for their potential on enabling diverse applications, such as secure optical communications \cite{Wu:13, Xue:16, Li:17}, high speed random bit generators \cite{Argyris:16, Li:16, Ugajin:17}, chaotic radar \cite{1303798}, and photonic spiking neural networks \cite{8693533}.   

Time delayed optical feedback can be achieved from an external mirror \cite{Oliver:11, Friart:14}, from phase-conjugate feedback \cite{PhysRevA.63.033805, PhysRevA.84.043836, Malica:20}, or from optoelectronic feedback \cite{Ma:20, PhysRevE.103.042206}. As the feedback light is coupled back to the laser cavity it alters drastically the laser output. The variations of the laser's output intensity dynamics are investigated numerically based on a set of delayed differential equations and the resulting dynamics depend on several feedback parameters, such as the optical phase, polarization, length of the external cavity, and the feedback rate.  

Recently, fiber Bragg gratings (FBGs) have been considered to provide time delayed feedback. An FBG differs from a mirror because it provides spatially distributed reflections along the length of the fiber which means that the feedback is also distributed in time. It has been shown that FBG feedback triggers rich dynamics, from stable to period-one oscillatory, quasi-periodic pulsating, period-doubled oscillatory, and chaotic \cite{6320711}. The undesirable time-delay signature (TDS) arising in feedback induced chaos can be efficiently removed by using FBG feedback. This is attributed to the dispersion at frequencies near the edge of the main lobe of the FBG reflectivity spectrum \cite{7097638}. In other studies, modifications of the grating profile have been reported, such as employing a Gaussian apodized FBG \cite{jun-feng_characteristics_2017} or chirped FBG (CFBG) \cite{Wang:17, Wang:19} to replace conventional uniform FBGs. Compared with uniform grating feedback, CFBG feedback is a completely dispersive feedback, and can eliminate the TDS without necessary frequency detuning \cite{CHAO2020124702}.  

These theoretical and experimental results indicate that the quality of the generated chaos can be improved by dispersion-induced optical feedback \cite{7882637, Zhang:18}. However, the influence of the FBG parameters such as bandwidth, maximum reflectivity or length have not been investigated. In this sense, identifying the most suitable FBG parameter can be of great interest since it might allow for the optimization of the chaotic behavior and further suppression of the TDS. Moreover, for very long grating lengths the reflection profile would resemble the so-called long faint gratings \cite{10.1117/12.2059668} which represent interest for study as filtered optical feedback with particular features as well as since they are crucial component for long-distance distributed sensing \cite{Yang:22}.

Here, we study the impact that the distribution of feedback in time has on the dynamics of semiconductor lasers. By adjusting the FBG parameters, we can modify the time-distribution. In order to isolate the effect of the time distribution alone we consider gratings with different lengths, but with the same impulse response shape. The complexity of the dynamics is analyzed by computing the largest Lyapunov exponent and the system's stability by tracking the position of the first Hopf bifurcation. We report the emergence of stability fluctuations of the laser behavior when long, narrow-bandwidth gratings are used, and find the dependence of this behavior on the feedback and laser parameters. We also highlight that the stability fluctuations are impacted differently for positive and negative detuning between the Bragg wavelength and the emission wavelength of the laser.

The paper is organized as follows. We first discuss the interdependence of the FBG parameters and their influence on the feedback time distribution in Sec. \ref{sec:theory} and discuss the dimensionless rate equation model used. In Sec. \ref{sec:stability_oscillations}, we analyze the effects of feedback time-distribution on the laser stability. In Sec. \ref{sec:FeedbackPhase} we study the impact of the feedback phase. Next, in Sec. \ref{sec:detuning}, we investigate the effect of detuning between the laser and the Bragg peak of the FBG. Finally, we discuss our results and conclude in Sec. \ref{sec:conclusion}. 

\section{Theory and model}
\label{sec:theory}
\begin{figure}[b]
\begin{center}
\includegraphics[width=\linewidth]{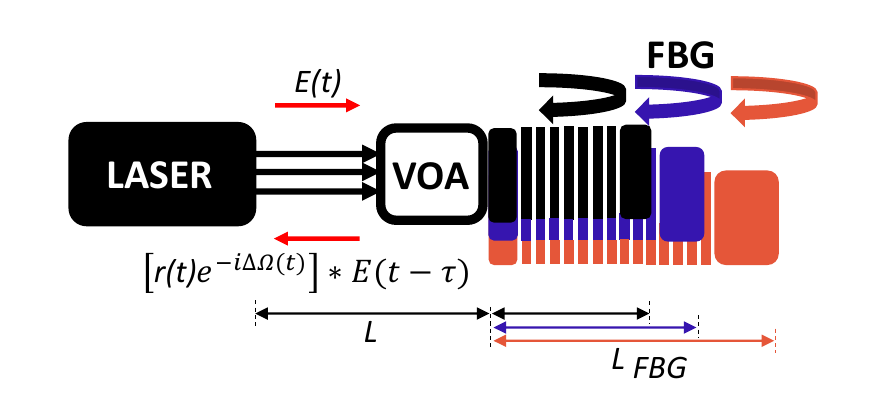}
\end{center}
\caption{\label{fig:setup} Schematic of a semiconductor laser subject to distributed feedback from a FBG, where $L$ is the distance between the laser output facet and the input of the FBG, which gives a feedback round-trip time of $\tau=2L/c$ with $c$ being the speed of light in vacuum. The variable optical attenuator (VOA) is set to change the feedback rate. $L_\textrm{FBG}$ is the length of the FBG which is varied, $E(t)$ is the field amplitude emitted by the laser, and $r(t)$ the impulse response of the FBG.}
\end{figure}

The schematic diagram of a semiconductor laser subjected to distributed feedback from an FBG is shown in FIG. \ref{fig:setup}. The change in the configuration from the replacement of a simple mirror by an FBG enables different reflection rates for different wavelengths according to the reflection spectrum of the grating. 

%setup schematics

The dynamics of the laser are simulated using a modified version of the single mode Lang-Kobayashi rate equations \cite{766841, 1070479}. The laser is described by the normalized intracavity optical field amplitude $E(t)$ and its normalized charge-carrier density $N(t)$. The rate equations governing the laser dynamics are \cite{6384650, 6320711}:
\begin{flalign}
\frac{dE}{dt} &= (1+i\alpha)NE+K e^{-i \Phi_0}\left[r(t)e^{-i\Delta\Omega t}\right]\ast E(t-\tau),
\label{eq:rate_equations1}\\
\frac{dN}{dt} &= \frac{1}{T}\left(P-N(t)-(1+2N(t))|E(t)|^2\right),
\label{eq:rate_equations2}
\end{flalign}
with:
\begin{multline}
r(\Omega) =\Omega_{BW}\\
\times\left(2\Omega +i\sqrt{\Omega_{BW}^2-4\Omega^2}\coth{\frac{\pi\sqrt{\Omega_{BW}^2-4\Omega^2}}{2\Omega_l}}\right)^{-1}.
\label{eq:impulse_response}   
\end{multline}

The equations are normalized in time by the photon lifetime and thus all time-related parameters are unitless. $T$ is the normalized carrier lifetime, $P$ is the pump parameter ($P=0$ corresponds to the threshold), and $\alpha$ the linewidth enhancement factor. In the simulations the values corresponding to $T, P$ and $\alpha$ are $1000, 1$ and $3$ respectively, unless stated otherwise. These are the dynamical parameters within the laser itself, while the FBG feedback is described by the last term in Eq. \ref{eq:rate_equations1}. Expressing the impulse response of the FBG reflection by $r(t)$, the field amplitude coupling back to the laser is proportional to the convolution $r(t)\ast E(t-\tau)$. $r(t)$ for the FBG case is obtained from the inverse Fourier transform of the FBG reflection frequency response $r(\Omega)$ given by Eq. \ref{eq:impulse_response} as reported in \cite{1233724, Li:14, 618322}. $K$ represents the normalized feedback rate. $e^{-i \Phi_0}$ is the total phase that the light accumulates while traveling in the feedback cavity where $\Phi_0=\omega_0\tau$ is the offset phase of the system with $\tau$ the external cavity round-trip time, and $\omega_0$ the dimensionless angular frequency of the free-running laser. $\Delta\Omega$ is the frequency detuning, defined as the difference of the Bragg frequency of the FBG with the free-running frequency of the laser. The FBG parameters are $\Omega_{BW}=2\pi f_{BW}$ and $\Omega_L=2\pi f_L$, where $f_{BW}$, and $f_L$ represent the full width at half maximum (FWHM) bandwidth of the main lobe in the reflectivity profile and the round-trip propagation time inside the FBG, respectively. Considering that the rate equations are normalized in time by the photon lifetime $\tau_p$, we  get the following relations:
\begin{align}
  \left(\frac{\Omega_L}{2\pi}\right) [\textrm{Hz}] &= \frac{\Omega_L [\textrm{a.u.}]}{1000 \tau_p [\textrm{s}]}
\label{eq:normalization1}\\
  \left(\frac{\Omega_{BW}}{2\pi}\right) [\textrm{Hz}] &= \frac{\Omega_{BW} [\textrm{a.u.}]}{1000 \tau_p [\textrm{s}]}
\label{eq:normalization2}\\
  L_{\textrm{FBG}} [\textrm{m}] &= \frac{c [\textrm{m}\cdot\textrm{s$^{-1}$}]}{2n_\textrm{eff} \left(\Omega_L/2\pi\right) [\textrm{Hz}]}
  \label{eq:normalization3}
\end{align}
In the simulations conducted here, we have chosen a photon lifetime value of $\tau_p = 1$\,ps for the sake of simplicity, although it should be noted that this is relatively low compared to the typical expected values of around $3$-$5$\,ps. The normalized $\Omega_{BW}$ and $\Omega_L$ together are related with the physical parameters according to the relation: 
\begin{equation}
  \kappa L_\textrm{FBG}=\frac{\pi\Omega_{BW}}{2\Omega_L}
\label{eq:FBGparameters}  
\end{equation}
where $\kappa$ is the coupling coefficient, typically expressed in units of $m^{-1}$, quantifying the strength of the dynamic coupling between the guided modes of the fiber and the fiber grating and $L_\textrm{FBG}$ is the length of the grating. 

%FBG parameters figure
\begin{figure}
\begin{center}
\includegraphics[width=\linewidth]{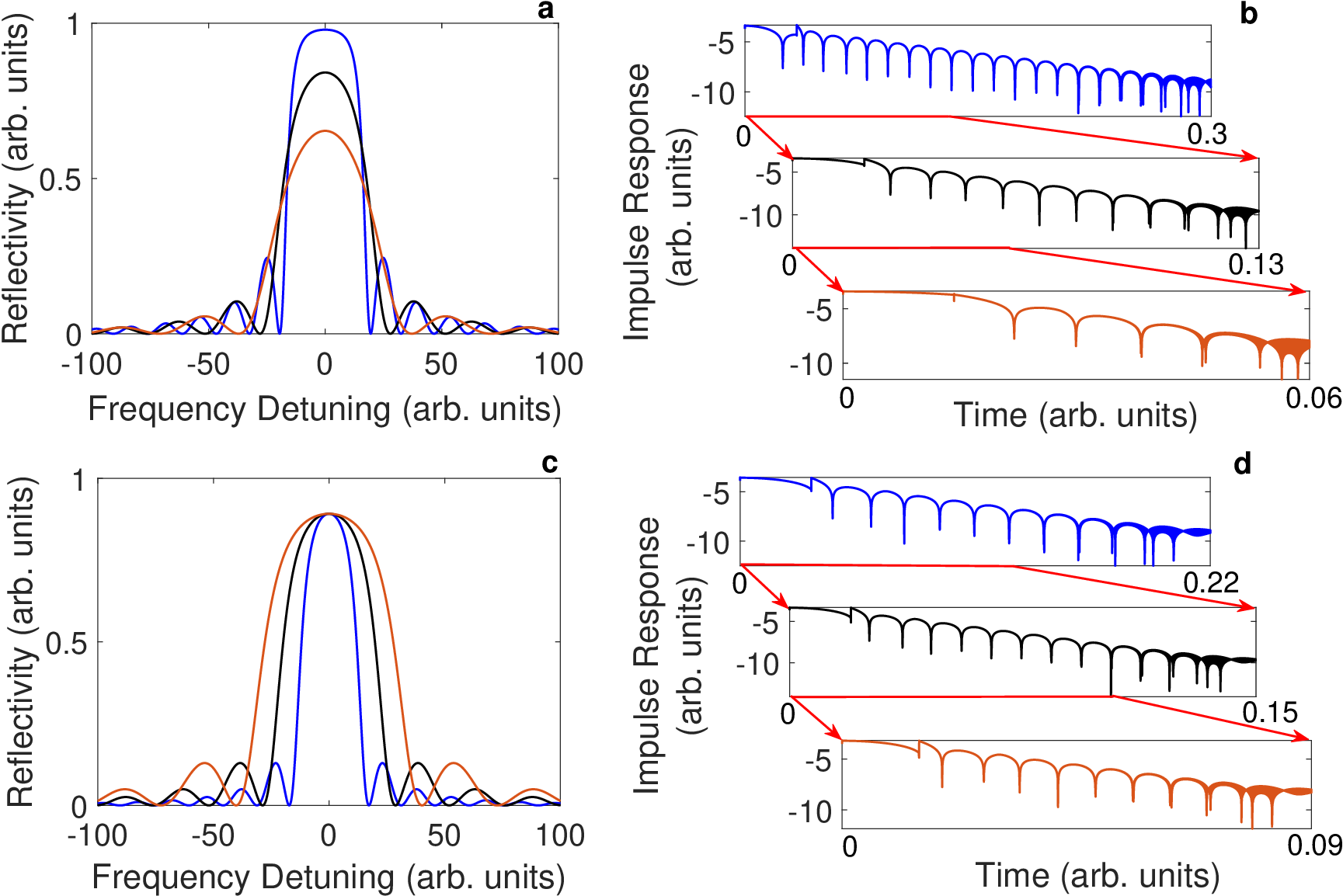}
\end{center}
\caption{\label{fig:FBG_parameters} (a) Reflection spectra of the FBG for varying grating length and maximum reflectivity with constant bandwidth; blue: $\Omega_L=15$, $\Omega_{BW}=25$; black: $\Omega_L=25$, $\Omega_{BW}=25$; orange: $\Omega_L=35$, $\Omega_{BW}=25$. (b) the impulse response for each of the cases in panel (a), plotted on a logarithmic scale on the y-axis. (c) Reflection spectra of the FBG varying the grating length and bandwidth to ensure a constant maximum reflectivity; blue: $\Omega_L=15$, $\Omega_{BW}=16.96$; black: $\Omega_L=25$, $\Omega_{BW}=28.27$; orange: $\Omega_L=35$, $\Omega_{BW}=39.57$; (d) the impulse response for each of the cases in panel (c), plotted on a logarithmic scale on the y-axis.}
\end{figure}

Here, we consider the interdependence between the grating parameters and their influence on the impulse response of the feedback. To vary the time distribution of the impulse response the grating length should change. %The latter is connected with the parameters in Eq. \ref{eq:FBGparameters} as $L_\textrm{FBG}\propto 1/\Omega_L$. 
The peak reflectivity of the FBG is attained at the Bragg frequency of $\Omega=0$ and is given by $\tanh^2(\kappa L_\textrm{FBG})$ . Thus, changing the length of the FBG will change the grating's maximum reflectivity and the shape of the reflectivity spectrum as illustrated in FIG. \ref{fig:FBG_parameters}(a). This means that the feedback signal input to the laser will change in terms of both time distribution and amplitude as shown in FIG. \ref{fig:FBG_parameters}(b). In order to guarantee the study of the effect of the time distribution alone, the shape of the impulse response should be fixed for different lengths. To attain this, we vary both $\Omega_{BW}$ and $\Omega_L$ such that the maximum reflectivity is unchanged but the bandwidth changes i.e., the $\kappa L_\textrm{FBG}$ product remains constant (FIG. \ref{fig:FBG_parameters}(c)). This will ensure that the shape of the impulse response remains the same when its distribution in time varies as shown in FIG. \ref{fig:FBG_parameters}(d). Consequently, in our simulation, whenever we change $\Omega_L$, we change $\Omega_{BW}$ according to Eq. \ref{eq:FBGparameters}.

The analysis of the system here is performed based on the calculations of the largest Lyapunov exponent (LLE) which is commonly used as a benchmark for chaos identification. The LLE gives the rate of separation of infinitesimally close trajectories in phase space to distinguish different attractors, and thus is useful in quantifying the complexity of the system's behavior \cite{doi:https://doi.org/10.1002/3527604804.ch6, ECKHARDT1993100}. Our focus has been on studying the stability of the system for different time-distributed feedback, so for that we track the transition between stable states to periodic oscillations by counting the number of extrema on the generated time series. 

\section{Different feedback time-distribution leads to stability fluctuations}
\label{sec:stability_oscillations}
\begin{figure}
\begin{center}
\includegraphics[width=\linewidth]{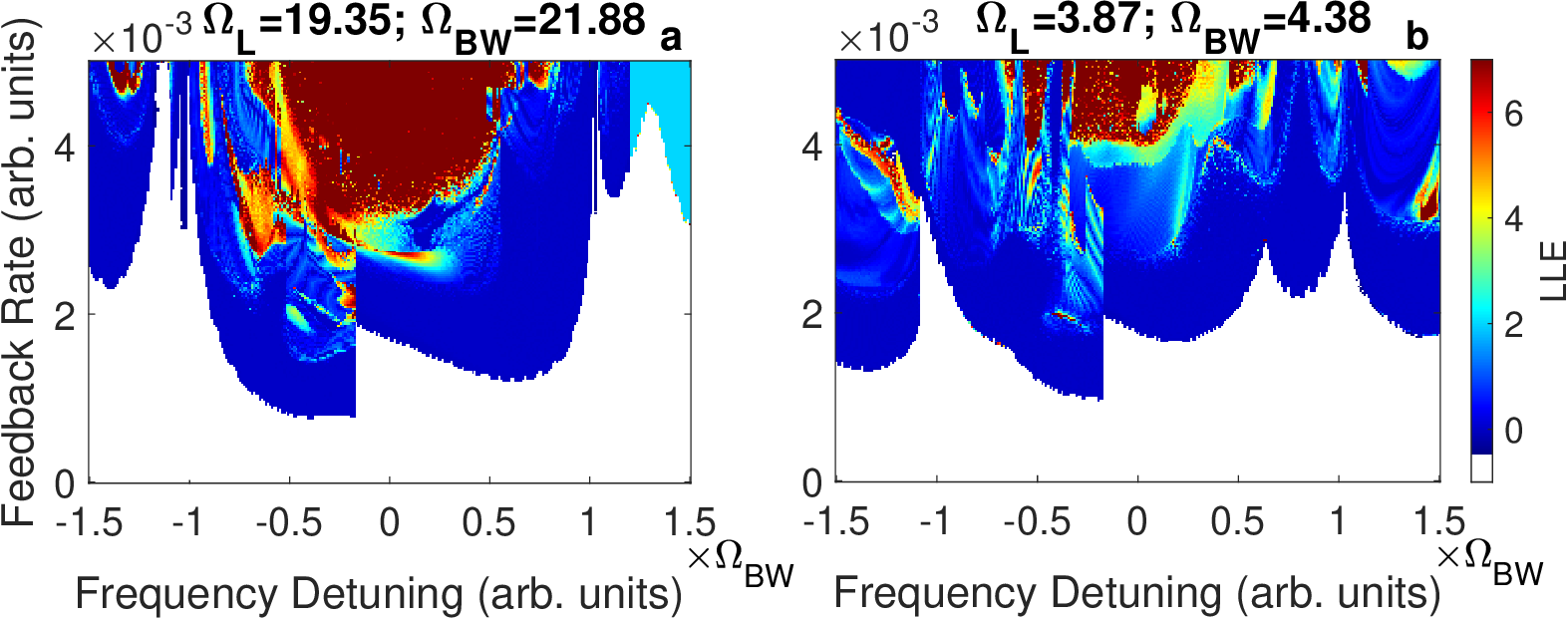}
\end{center}
\caption{\label{fig:DifferentLengths} Numerical mapping of the values of the LLE in the parameter space of feedback rate $K$ and grating detuning frequency $\Delta\Omega$ for (a) $\Omega_L=19.35$ and $\Omega_{BW}=21.88$ ($L_\textrm{FBG}=0.5$\,cm) and (b) $\Omega_L=3.87$ and $\Omega_{BW}=4.38$ ($L_\textrm{FBG}=2.5$\,cm). For both gratings the value of the maximum reflectivity, $\kappa L_\textrm{FBG}$ is $1.776$.}
\end{figure}

 \begin{table*}
\caption{\label{tab:oscfreq}The values of $L_\textrm{FBG}$ where the $n^{th}$ zero overlaps with the laser side lobe around the relaxation oscillation, extracted from the simulation, and predicted by the hypothesis and their difference $\Delta$ for different values of $\kappa L_\textrm{FBG}$ at zero detuning.}  
%\texttt{table} 
%\textbackslash\texttt{multicolumn}
\begin{ruledtabular}
\begin{tabular}{{c|ccc|ccc|ccc}} 
\rule[-1ex]{0pt}{3ex} &\multicolumn{9}{c}{Peaks position (cm)} \\
\hline
\rule[-1ex]{0pt}{3ex}  $\kappa L_\textrm{FBG}$ & \multicolumn{3}{c}{0.888} & \multicolumn{3}{c}{1.776} &\multicolumn{3}{c}{3.552}  \\
\hline
\rule[-1ex]{0pt}{3ex}  n & Hypth.  & Sim. & $\Delta$ & Hypth. & Sim. & $\Delta$ & Hypth. & Sim. & $\Delta$  \\
\hline
\rule[-1ex]{0pt}{3ex}  1 & 1.51 & 1.50 & 0.01 & 1.67 & 1.64 & 0.03 & 2.19 & 2.09 & 0.10  \\
\rule[-1ex]{0pt}{3ex}  2 & 2.93 & 2.91 & 0.02 & 3.02 & 2.98 & 0.04 & 3.34 & 3.17 & 0.17 \\
\rule[-1ex]{0pt}{3ex}  3 & 4.38 & 4.33 & 0.05 & 4.44 & 4.35 & 0.09 & 4.66 & 4.44 & 0.22  \\
\rule[-1ex]{0pt}{3ex}  4 & 5.84 & 5.71 & 0.13 & 5.87 & 5.69 & 0.18 & 6.04 & 5.74 & 0.30 \\
\rule[-1ex]{0pt}{3ex}  5 & 7.28 & 7.08 & 0.20 & 7.32 & 7.08 & 0.24 & 7.46 & 7.13 & 0.33  \\
\end{tabular}
\end{ruledtabular}
\end{table*}

In FIG. \ref{fig:DifferentLengths} we show a numerical map of the values of the LLE for two gratings with different lengths in the feedback parameter space of grating detuning frequency, $\Delta\Omega$, and feedback strength, $K$. Since the LLE can be used to quantify the chaos in a system, we can easily observe the evolution of the laser dynamics. The white regions, for LLE $<0$, represents the regions where the laser is in a stable state. While the most complex chaotic behavior, for high values of the LLE, is displayed in red. The color gradient in-between represents the gradual increase of the LLE between the two extreme cases. 
There is a correlation between the feedback strength required for instability and the FBG reflectivity spectrum, thus, the stabilization regions result from the dips of the FBG and the periodicity resembles that of the side lobes of the FBG reflection spectrum. Though the evolution of the dynamics in the ($K,\Delta\Omega$) space follows the inverse of the FBG spectrum, its evolution for different grating lengths varies. Specifically the dynamics are observed at different feedback rates as the frequency detuning changes. 

\begin{figure}[ht]
\begin{center}
\includegraphics[width=\linewidth]{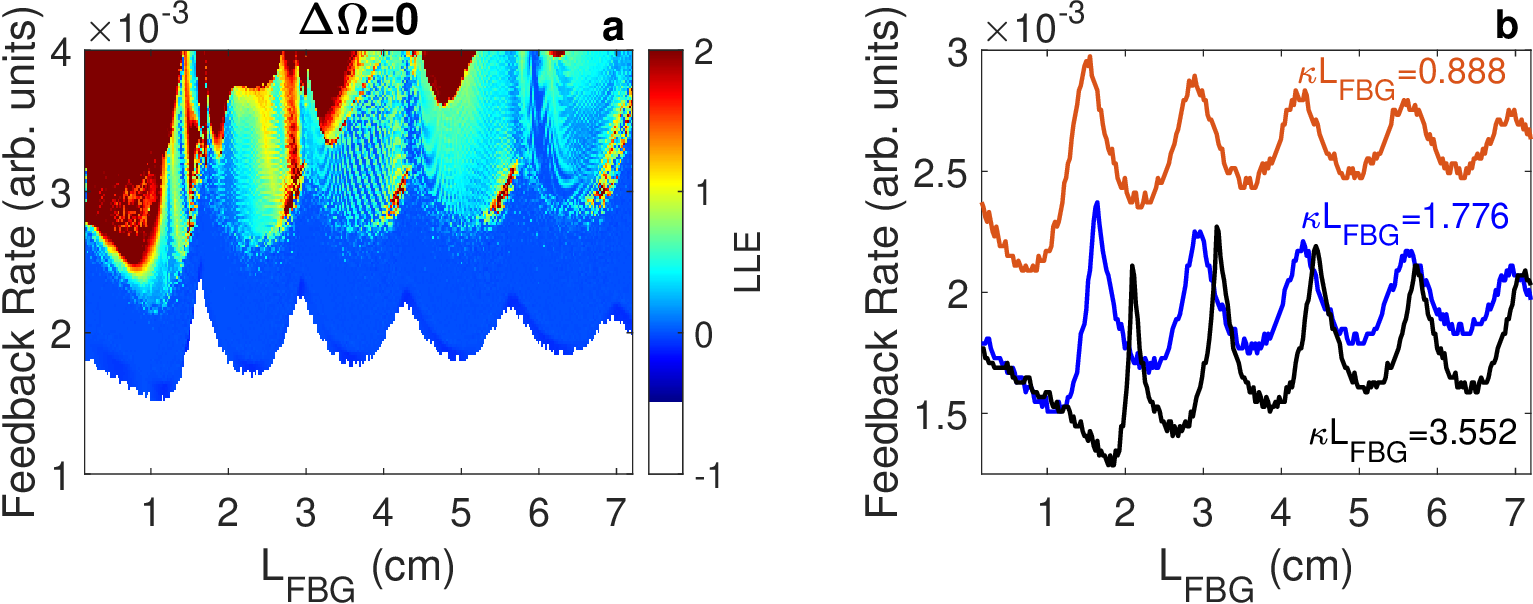}
\end{center}
\caption{\label{fig:PeriodicOscillations}(a) Numerical map of the values of the LLE as a function of $L_\textrm{FBG}$ and feedback rate for a grating with $\kappa L_\textrm{FBG}=1.776$ at zero detuning. (b) Fluctuations of the laser’s stability observed for $L_\textrm{FBG}>1$\,cm by tracking the occurrence of the Hopf bifurcation while switching the feedback ratio for $\kappa L_\textrm{FBG}=0.888$ (orange), $\kappa L_\textrm{FBG}=1.776$ (blue), and $\kappa L_\textrm{FBG}=3.552$ (black).}
\end{figure}

To explore the laser dynamics for different grating lengths, we calculate the LLE in the parameter space of the grating length, $L_\textrm{FBG}$, and feedback rate for $\kappa L_\textrm{FBG}=1.776$ at zero detuning, as depicted in FIG. \ref{fig:PeriodicOscillations}(a). By selecting $\kappa L_\textrm{FBG}=1.776$, we achieve a reflectivity spectrum resembling real FBGs with maximum reflectivity of approximately $90$\%. The variations in the LLE and color gradient, detailed in the colorbar, help classify the dynamical states. Interestingly, certain grating lengths exhibit higher feedback rates required to destabilize the laser, leading to stability fluctuations. This trend persists during transitions between more complex dynamical states at higher feedback rates. To better understand this behavior, we analyze the evolution of the Hopf bifurcation, where the feedback ratio switches the system's state from stable to a periodic solution, as shown in blue in FIG. \ref{fig:PeriodicOscillations}(b). We observe that the stability boundary remains non-oscillatory for lengths below $\sim 1$\,cm, while stability fluctuations appear and repeat for longer gratings.

To examine the oscillatory behavior for different FBG maximum reflectivity values, we repeat the procedure for various $\kappa L_\textrm{FBG}$ values. FIG. \ref{fig:PeriodicOscillations}(b) shows the evolution of the Hopf bifurcation for $\kappa L_\textrm{FBG}=0.888$ (orange) and $\kappa L_\textrm{FBG}=3.552$ (black). We find that stability fluctuations are present for different reflectivity values, and overall, it is easier to destabilize the laser with larger $\kappa L_\textrm{FBG}$ values due to the higher reflectivity. The translation of peaks to higher grating lengths with increasing $\kappa L_\textrm{FBG}$ can be attributed to variations in the grating bandwidth. Additionally, we observe that the oscillation peaks become sharper with increasing $\kappa L_\textrm{FBG}$ due to the reflection profile's saturation and steep slope near the zeros of the reflectivity spectrum. Throughout the paper, we maintain a constant maximum reflectivity of $\kappa L_\textrm{FBG}=1.776$.

\begin{figure}[ht]
\begin{center}
\includegraphics[width=0.8\linewidth]{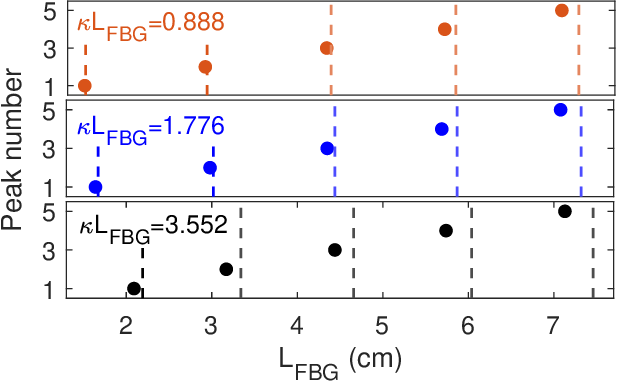}
\end{center}
\caption{\label{fig:Peaks_hypothesis} Position of the stability peaks at zero detuning extracted from the simulation for $\kappa L_\textrm{FBG}=0.888$ (orange), $\kappa L_\textrm{FBG}=1.776$ (blue), and $\kappa L_\textrm{FBG}=3.552$ (black) compared with the positions estimated from the hypothesis (dotted lines) for each case.}
\end{figure}

The existence of stability fluctuations in our system can be attributed to the interplay of time scales characterizing the filtered optical feedback (FOF) dynamics. Fischer et al. \cite{1366365} identified the relevant time scales as the laser's relaxation oscillation frequency $\Omega_{RO}$, the free spectral range of the external cavity mode, and the inverse of the filter bandwidth. As the feedback rate increases, the undamped relaxation oscillations dominate the laser dynamics, affecting its stability. Combining the relaxation oscillations with the specific shape of the FBG reflectivity spectrum explains the observed stability fluctuations. These stabilization peaks occur at specific bandwidths when the laser side lobe near the relaxation oscillation frequency overlaps with the zeros of the FBG reflectivity spectrum. 
Using the notation of \cite{618322} the position of the $n^{th}$ zero of reflectivity spectrum of the FBG is:
\begin{equation}
  \delta_n=\sqrt{\kappa^2+\left(n\frac{\pi}{L_\textrm{FBG}}\right)^2}
\label{eq:deltan}  
\end{equation}
We then rewrite this equation to get the frequency at which the $n^{th}$ zero occurs:
\begin{equation}
     \delta f_n = \frac{\pi c}{L_\textrm{FBG} n_\textrm{eff}}\sqrt{n^2+\left(\frac{\kappa L_\textrm{FBG}}{\pi}\right)^2}
\label{eq:deltafn}   
\end{equation}
where $n_\textrm{eff}=1.5$ is the effective refractive index of the grating.
We employ Eq. \ref{eq:deltafn} to calculate the frequencies at which these zeros occur, and by comparing with the simulation results as shown in FIG. \ref{fig:Peaks_hypothesis}, we verify our statement (see TABLE \ref{tab:oscfreq} for details). The observed stability fluctuations arise from the overlap between the laser side lobe around the relaxation oscillation frequency and the zeros of the FBG reflectivity spectrum, or equivalently, the dips in the time distribution as shown in FIG. \ref{fig:FBG_parameters}(d). When the first minimum in the FBG spectrum coincides with the relaxation oscillation frequency, the first minimum in the temporal response corresponds to the relaxation oscillation period. Such a temporal response means that the oscillations in time are severely damped by the FBG, agreeing with the spectral picture that we use here.

\section{Progression of stability fluctuations with the feedback offset phase}
\label{sec:FeedbackPhase}
The laser's emission dynamics also depend heavily on the feedback phase, that is, the phase that the light accumulates while traveling through the feedback loop. In the case of filtered optical feedback, where the filter alters the spectral content of the feedback light, lasers show different types of dynamics and dependence on the feedback phase \cite{PhysRevE.76.026212}. Recently, we have shown that semiconductor lasers have a strong sensitivity to the phase of time-distributed feedback from an FBG \cite{Skenderas:22}. Specifically, the offset phase of the system changes the overall dynamics exhibited by the laser influencing this way also the periodic stability variations in the border between different dynamical states. Depending on the initial conditions these variations could fade or become more prominent. In this regard, it is of interest to study the effects of the offset phase variations on the stability fluctuations that emerge for different feedback time-distributions explained above. 
%offset Phase and Peaks
\begin{figure}[ht]
\begin{center}
\includegraphics[width=0.95\linewidth]{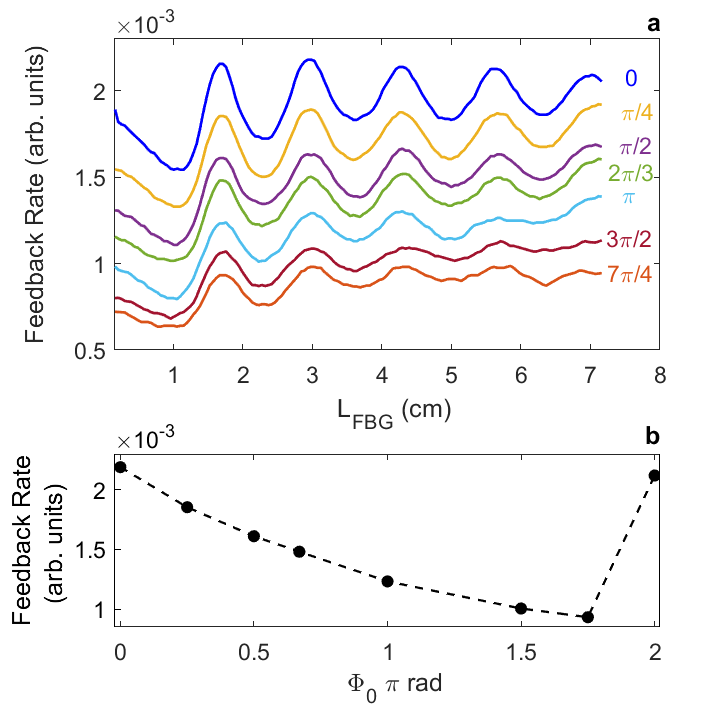}
\end{center}
\caption{\label{fig:offsetPhaseAndPeaks} (a) The evolution of Hopf bifurcation tracking the stability fluctuations as function of $L_\textrm{FBG}$ at zero detuning for different values of the feedback offset phase $\Phi_0$ equal to $0$ (blue), $\pi/4$ (yellow), $\pi/2$ (violet), $2\pi/3$ (green), $\pi$ (cyan), $3\pi/2$ (maroon), and $7\pi/4$ (orange). (b) The feedback rate necessary to destabilize the laser as a function of the offset phase $\Phi_0$ at zero detuning for the first stabilization peak located at $L_\textrm{FBG}=1.69$\,cm.}
\end{figure}

%Detuning figure
\begin{figure*}
\begin{center}
\includegraphics[width=0.98\linewidth]{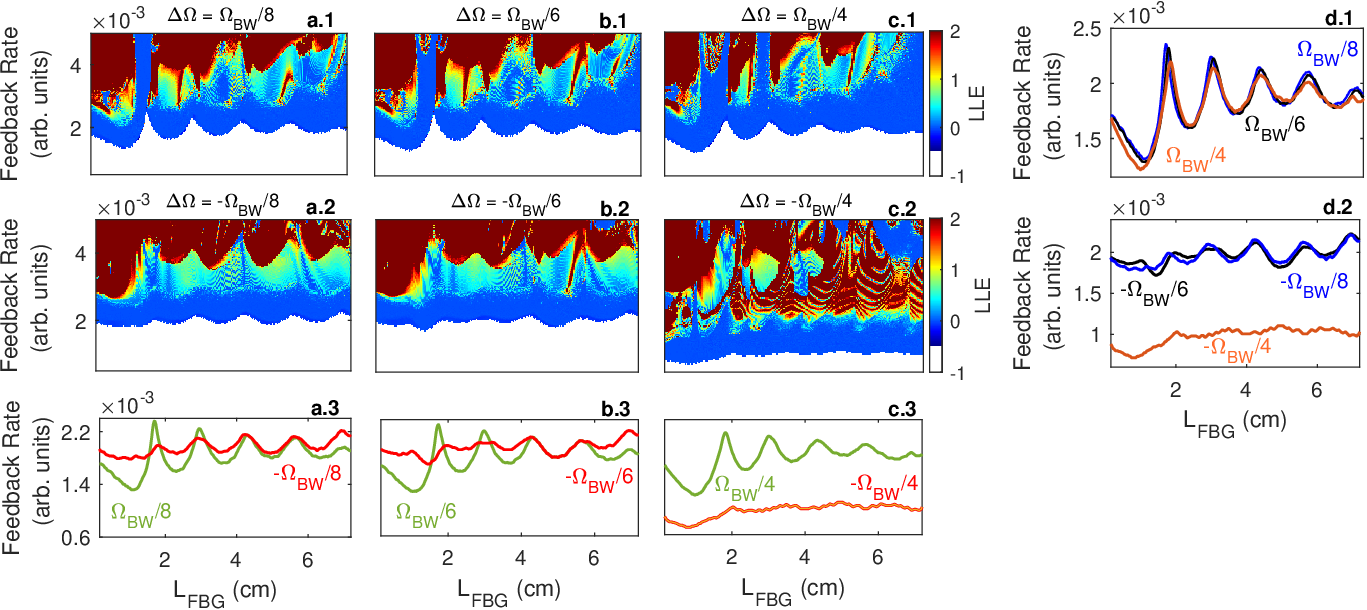}
\end{center}
\caption{\label{fig:Detuning} Numerical maps of the values of the largest Lyapunov exponent as a function of $L_\textrm{FBG}$ and feedback rate for a grating with $\kappa L_\textrm{FBG}=1.776$ and positive detuning $\Omega_{BW}/8$ (a.1), $\Omega_{BW}/6$ (b.1), $\Omega_{BW}/4$ (c.1) and negative detuning -$\Omega_{BW}/8$ (a.2), -$\Omega_{BW}/6$ (b.2), -$\Omega_{BW}/4$ (c.2). The evolution of the Hopf bifurcation for the three positive detuning (d.1) and the three negative detuning (d.2). The comparison of the occurrence of the Hopf bifurcation for $\Omega_{BW}/8$ and -$\Omega_{BW}/8$ (a.3),  $\Omega_{BW}/6$ and -$\Omega_{BW}/6$ (b.3), $\Omega_{BW}/4$ and -$\Omega_{BW}/4$ (c.3).}
\end{figure*}

To investigate the impact that the feedback offset phase, $\Phi_0$, has on the stability fluctuations we explore the progression of the laser dynamics while varying the grating length for eight different $\Phi_0$ at zero detuning (which means the total phase is also $\Phi_0$). Again, the stability fluctuations are present for each case and the same variations of the dynamics are present. The progression of the Hopf bifurcation for each map is shown in FIG. \ref{fig:offsetPhaseAndPeaks}(a) for the different $\Phi_0$. We can observe that as the offset phase increases, the required feedback rate to destabilize the laser decreases and the oscillations fade i.e., the border between the stable and periodic behavior becomes smoother until the sudden change for $\Phi_0=2\pi$.
We should note that the position of the stability peaks is not altered by the variations of the offset phase which consequently strengthen the argument that the fluctuations in stability are due to the interdependence of the zeros of the FBG reflection spectrum and the side lobes around the relaxation oscillation frequency, and the variation of the feedback phase only changes the feedback rate at which the stability peaks appear. The variation of the feedback rate of the stability fluctuations with the $\Phi_0$ for the first stabilization peak located at $L_\textrm{FBG}=1.69$\,cm is shown in FIG. \ref{fig:offsetPhaseAndPeaks}(b) with the black dotted line. The feedback rate at which the stabilization peak appears decreases almost linearly until $\Phi_0 = 7\pi/4$, and then starts to increase again until it reaches its initial value for $\Phi_0 = 2\pi$. 

\section{FBG detuning and stability fluctuations}
\label{sec:detuning}

\begin{figure} %cartoon detuning
\begin{center}
\includegraphics[height=5.4cm]{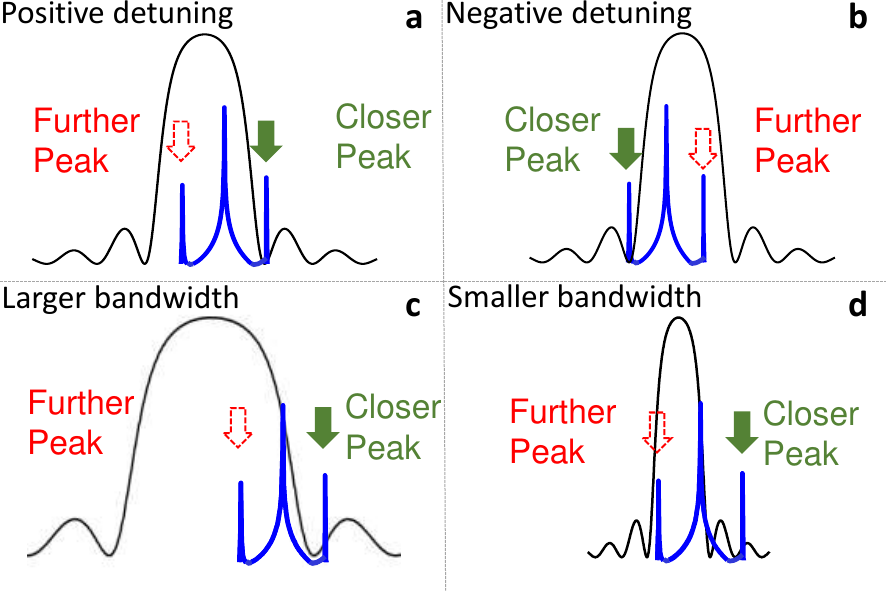}
\end{center}
\caption{\label{fig:DetuningSchematics} Schematic representation to illustrate the relative position of the zeros of the zeros of the FBG reflectivity and the side lobes of the laser's spectrum: (a) for positive detunings, (b) for negative detunings, and for the variation of the grating's bandwidth (c) large bandwidth, (d) smaller bandwidth.}
\end{figure}

It has been shown that different frequency detuning $\Delta\Omega$ can lead to distinct dynamical states e.g. the TDS suppression prefers the FBG at a positive frequency detuning \cite{7097638}. Hence, it is interesting to investigate the progression of the observed stability fluctuations for various frequency detuning. To quantify the changes we calculate again the LLE for six different detunings, three positive and three negative, symmetric with the filter's center frequency, the Bragg frequency. These detunings are always chosen as a fraction of the FBG bandwidth, $\Omega_{BW}$, in order to remove the effect of the reflectivity. This means that the absolute value of the detuning decreases with increasing $L_\textrm{FBG}$.
The numerical maps of the LLE for the positive detuning $\Omega_{BW}/8$, $\Omega_{BW}/6$, and $\Omega_{BW}/4$ are shown in FIG. \ref{fig:Detuning}(a.1), (b.1), and (c.1) respectively and for the negative detuning -$\Omega_{BW}/8$, -$\Omega_{BW}/6$, and -$\Omega_{BW}/4$ are shown in FIG. \ref{fig:Detuning}(a.2), (b.2), and (c.2) respectively. To investigate the stability of the laser at different frequency detuning we also plot and compare the evolution of the Hopf bifurcation.
We observe that the stability fluctuations, discussed previously, are present and with specific features for each case. In general, each map shows the typical route-to-chaos for the laser with frequency-detuned FBG feedback.

We examine the position of the stabilization peaks in relation to the laser side lobe around the relaxation oscillations. When a detuned FBG is used with a detuning value of $\Delta\Omega$, Eq. \ref{eq:deltafn} can be expressed as:
\begin{equation}
  L_\textrm{FBG}^{(\pm n)}=\frac{\pi c \left(\sqrt{n^2+\left(\frac{\kappa L_\textrm{FBG}}{\pi}\right)^2}\right)}{n_\textrm{eff}\left(\Omega_{RO}\pm\Delta\Omega\right)}
\label{eq:L_detuned}  
\end{equation}
where $\Omega_{RO}$ represents the frequency of the laser's relaxation oscillations. This results in two solutions for a given detuning and peak number $n$, which are equal to the two solutions with opposite detuning. When the detuning is non-zero, the zeros in the reflectivity profile are no longer symmetric to the laser's sidebands as shown in FIG. \ref{fig:DetuningSchematics}(a) and (b) for positive and negative detuning, respectively. Therefore, there exist two solutions, one for each of the laser's sideband positions. When the grating length is varied, the bandwidth is also changing to ensure the same maximum reflectivity, as explained, resulting in a change in the relative position of the side peaks of the laser around the relaxation oscillation frequency and the zeros of the FBG's reflectivity spectrum. This is illustrated in FIG. \ref{fig:DetuningSchematics}(c) and (d). As the detuning is altered, the zeros in the reflectivity spectrum move closer to one of the laser's sidebands, which we refer to as the closer peak, and further away from the other, which we call the further peak. 

\begin{figure}
\begin{center}
\includegraphics[width=0.65\linewidth]{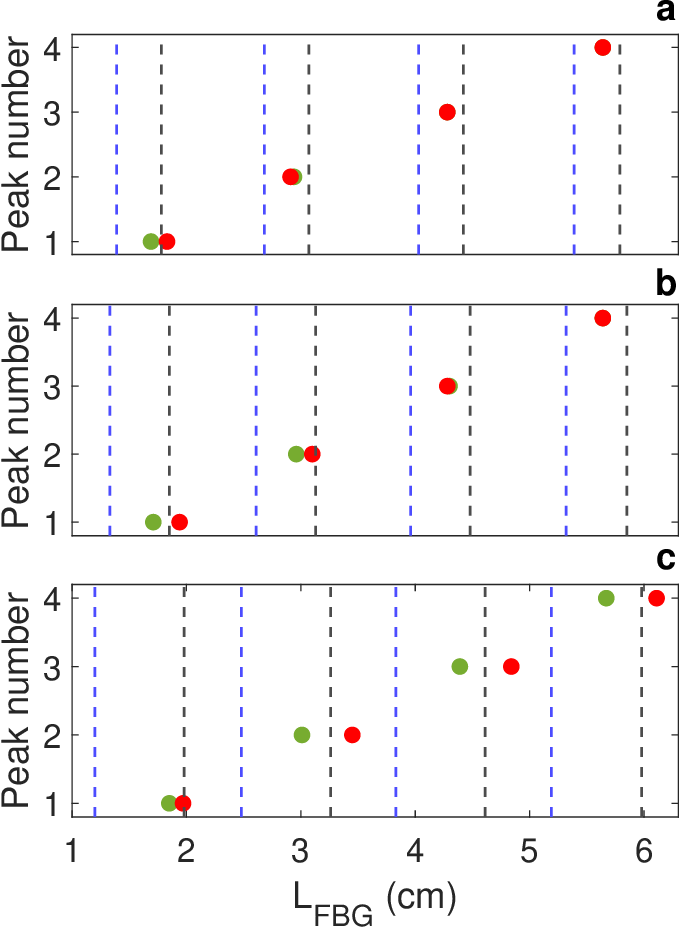}
\end{center}
\caption{\label{fig:Peaks_hypothesis2} Position of the simulated stability peaks shown with green and red dots for positive and negative detunings, correspondingly, compared with the theoretical positions in dashed blue and black lines for the closer and further peak respectively for $\Omega_{BW}/8$ (a), $\Omega_{BW}/6$ (b), and $\Omega_{BW}/4$ (c).} 
\end{figure}

The peak positions were determined using Eq. \ref{eq:L_detuned} and are represented by blue and black dotted lines for the closer and further peaks, respectively, in FIG. \ref{fig:Peaks_hypothesis2}. The green and red dots indicate the peak positions obtained from the simulation, i.e. those shown in FIG. \ref{fig:Detuning}, for positive and negative detuning, respectively. These results are summarized in TABLE \ref{tab:detuning}. 
Our analysis agrees well with the simulation results obtained for detuning $\pm\Omega_{BW}/8$ and $\pm\Omega_{BW}/6$, as shown in FIG. \ref{fig:Peaks_hypothesis2}(a) and (b), where the zeros of the FBG overlap with the further peak. 
\begin{table*}
\caption{\label{tab:detuning}The values of $L_\textrm{FBG}$ for which the $n^{th}$ zero overlaps with the laser side lobes around the relaxation oscillation frequency for different frequency detuning $\Delta\Omega$ corresponding to the maps in FIG. \ref{fig:Detuning}.}  
\begin{ruledtabular}
\begin{tabular}{{c|ccc|ccc|ccc}} 
\rule[-1ex]{0pt}{3ex} &\multicolumn{9}{c}{Peaks position (cm)} \\
\hline
\rule[-1ex]{0pt}{3ex}  $\Delta\Omega$ & \multicolumn{3}{c|}{$\Omega_{BW}/8$} & \multicolumn{3}{c|}{$\Omega_{BW}/6$} &\multicolumn{3}{c}{$\Omega_{BW}/4$}  \\
\hline
\rule[-1ex]{0pt}{3ex}  n & Hypth.  & Sim. &$\Delta$ & Hypth. & Sim. &$\Delta$ & Hypth. & Sim. &$\Delta$   \\
\hline
\rule[-1ex]{0pt}{3ex}  1-Right & 1.39 & 1.69 & 0.30 & 1.33 & 1.71 & 0.38 & 1.20 & 1.85 & 0.65  \\
\rule[-1ex]{0pt}{3ex}  2-Right & 2.68 & 2.94 & 0.26 & 2.61 & 2.96 & 0.35 & 2.48 & 3.01 & 0.53  \\
\rule[-1ex]{0pt}{3ex}  3-Right & 4.03 & 4.28 & 0.25 & 3.96 & 4.30 & 0.34 & 3.83 & 4.39 & 0.56  \\
\rule[-1ex]{0pt}{3ex}  4-Right & 5.39 & 5.64 & 0.25 & 5.32 & 5.64 & 0.32 & 5.19 &  5.67 & 0.48 \\
\hline
\rule[-1ex]{0pt}{3ex}  $\Delta\Omega$ & \multicolumn{3}{c|}{-$\Omega_{BW}/8$} & \multicolumn{3}{c|}{-$\Omega_{BW}/6$} &\multicolumn{3}{c}{-$\Omega_{BW}/4$}  \\
\hline
\rule[-1ex]{0pt}{3ex}  n & Hypth.  & Sim. &$\Delta$ & Hypth. & Sim. &$\Delta$ & Hypth. & Sim. &$\Delta$   \\
\hline
\rule[-1ex]{0pt}{3ex}  1-Left & 1.78 & 1.83 & 0.05 & 1.85 & 1.94 & 0.09 & 1.98 & 1.97 & 0.01  \\
\rule[-1ex]{0pt}{3ex}  2-Left & 3.07 & 2.91 & 0.16 & 3.13 & 3.10 & 0.03 & 3.26 & 3.45 & 0.19 \\
\rule[-1ex]{0pt}{3ex}  3-Left & 4.42 & 4.28 & 0.14 & 4.48 & 4.28 & 0.20 & 4.61 & 4.84 & 0.23  \\
\rule[-1ex]{0pt}{3ex}  4-Left & 5.79 & 5.64 & 0.15 & 5.85 & 5.64 & 0.21 & 5.98 & 6.11 & 0.13  \\
\end{tabular}
\end{ruledtabular}
\end{table*}

Interestingly, whether the frequency detuning is positive or negative, the resulting stabilization peaks always originate from the position of the further peak. As the value of $L_\textrm{FBG}$ increases, the difference between the simulated and calculated positions of the stabilization peaks becomes more pronounced. This can be attributed to the varying relative positions of the laser's side peaks around the relaxation oscillation frequency and the zeros of the FBG's reflectivity spectrum, as illustrated in FIG. \ref{fig:DetuningSchematics}(c) and (d). The imperfect overlap between these two components leads to a greater discrepancy in the peak positions.
Specifically, for a detuning of -$\Omega_{BW}/4$, the differences become more pronounced and noticeable. The feedback rate at which destabilization occurs is lower, and the stabilization border appears flatter. This can be attributed to the variation in the FBG phase, as shown in FIG. \ref{fig:DifferentLengths}, which exhibits a sharp transition at around this detuning. However, despite these differences, the simulation values for the stabilization peak in both detunings, $\Omega_{BW}/4$ and -$\Omega_{BW}/4$, are close to the further peak. The deviation, although higher, may arise from the intermediate position of the further peak, as depicted in FIG. \ref{fig:DetuningSchematics}(c) and (d).

Furthermore, we have observed an asymmetry in the stability regions with respect to detuning, as shown in FIG. \ref{fig:Detuning}(a.3), (b.3), and (c.3). This is a unique characteristic of FBG feedback that has not been reported before in filtered optical feedback systems. In the remaining part of this section, we delve into several aspects of the FBG feedback that could explain the observed asymmetric behavior. We examine the feedback parameters specific to the FBG case, as well as the laser parameters that might influence the stability fluctuations observed earlier. 
Specifically, we consider the influence of the relaxation oscillation frequency since when subjected to the nonlinear response of the filter with varying feedback parameters the relaxation frequency of the laser changes \cite{uchida2012optical}. Additionally, we analyze the FBG phase, since it has a specific behavior which varies with the detuning, and the effective phase which alters the emission frequency of the laser.

\subsection{Variation of the relaxation oscillation frequency of the laser}
\label{sec:relaxation_oscillation}
\begin{figure} %peaks corrected for omega_RO
\begin{center}
\includegraphics[width=0.65\linewidth]{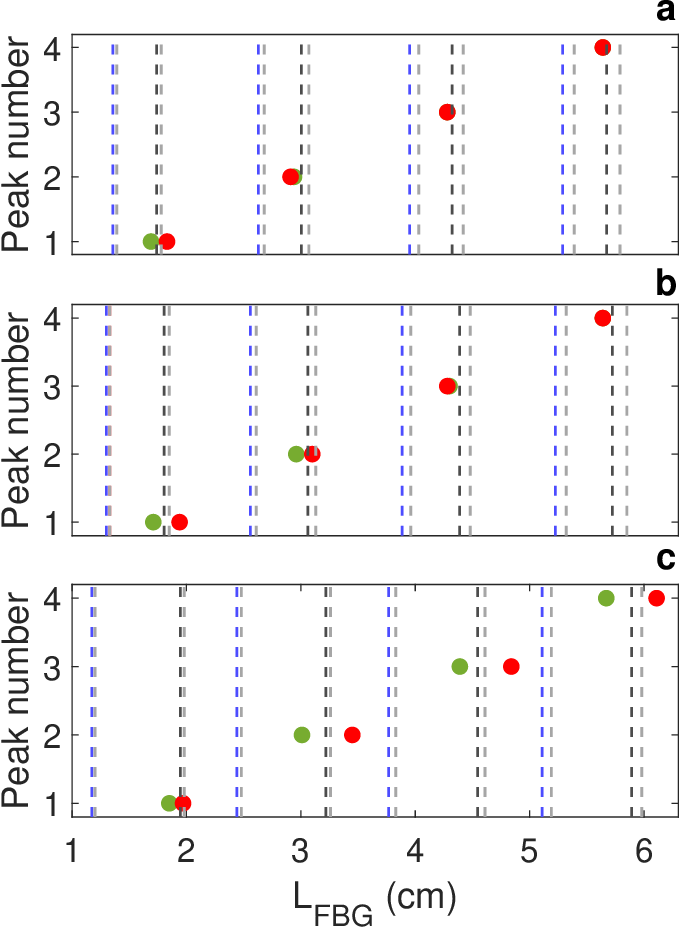}
\end{center}
\caption{\label{fig:Peaks_hypothesis2_RO} Position of the stability peaks extracted from the simulation  shown with green dots for positive detunings and red dots for the negative detunings compared with the positions estimated from the hypothesis for the corrected value of the relaxation oscillation frequency of the laser in dashed blue and black lines for the closer and further respectively for $\Omega_{BW}/8$ (a), $\Omega_{BW}/6$ (b), and $\Omega_{BW}/4$ (c). The corresponding values before correcting for the change in the relaxation oscillation frequency are shown with gray dotted lines.} 
\end{figure}
It has been reported that the relaxation oscillation frequency of the laser depends on the operation parameters of the external optical feedback, e.g., feedback strength, external cavity length \cite{PhysRevLett.65.1999, 8533, 119502} and feedback optical phase \cite{6459532, Liu:21}. From Eq. \ref{eq:L_detuned} and \ref{eq:L_detuned_plus_delta_nu} for alternations of the $\Omega_{RO}$ the positions of the stability peaks will change. Thus, for each combination of the feedback parameters at the stabilization peaks we calculate the resulting relaxation oscillation frequency from the Lang-Kobayashi equations and then re-estimate the theoretical grating length where the stabilization occurs, i.e., the position of the closer and further peak. The results obtained after correction for the variations of the relaxation oscillation frequency are shown in FIG. \ref{fig:Peaks_hypothesis2_RO} with dashed blue and black lines for the closer and further peak respectively, and overlapped with the simulation values of the stabilization peaks for positive (green dots) and negative (red dots) detuning.
The position of the dashed lines has changed slightly, as indicated by the grey dotted lines in FIG. \ref{fig:Peaks_hypothesis2_RO}, resulting in a better match with the simulation values for the further peak for detunings of $\pm\Omega_{BW}/8$ and $\pm\Omega_{BW}/6$, even for longer $L_\textrm{FBG}$. However, for detunings of $\Omega_{BW}/4$ and -$\Omega_{BW}/4$, there is still a significant deviation between the simulated and theoretical values of the stabilization peak position, despite a clear preference for the further peak. 

\subsection{Impact of the FBG phase}
\label{sec:FBGphase}
\begin{figure}
\begin{center}
\includegraphics[height=2.7cm]{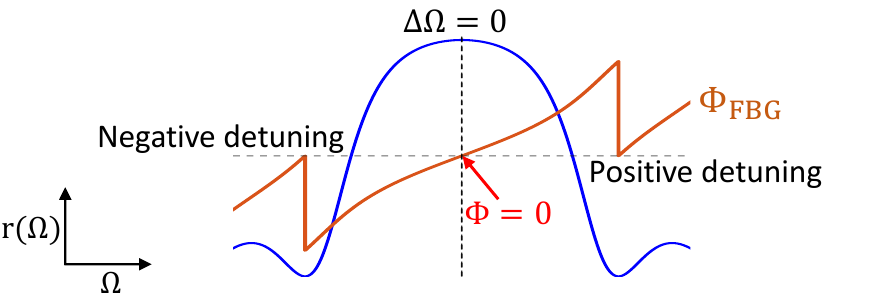}
\end{center}
\caption{\label{fig:FBGphaseSchematics} Schematic representation illustrating the variations of the FBG phase, $\Phi_\textrm{FBG}$, for different detuning along the FBG reflectivity spectrum. }
\end{figure}

%FBG phase figure
\begin{figure*}
\begin{center}
\includegraphics[width=0.9\linewidth]{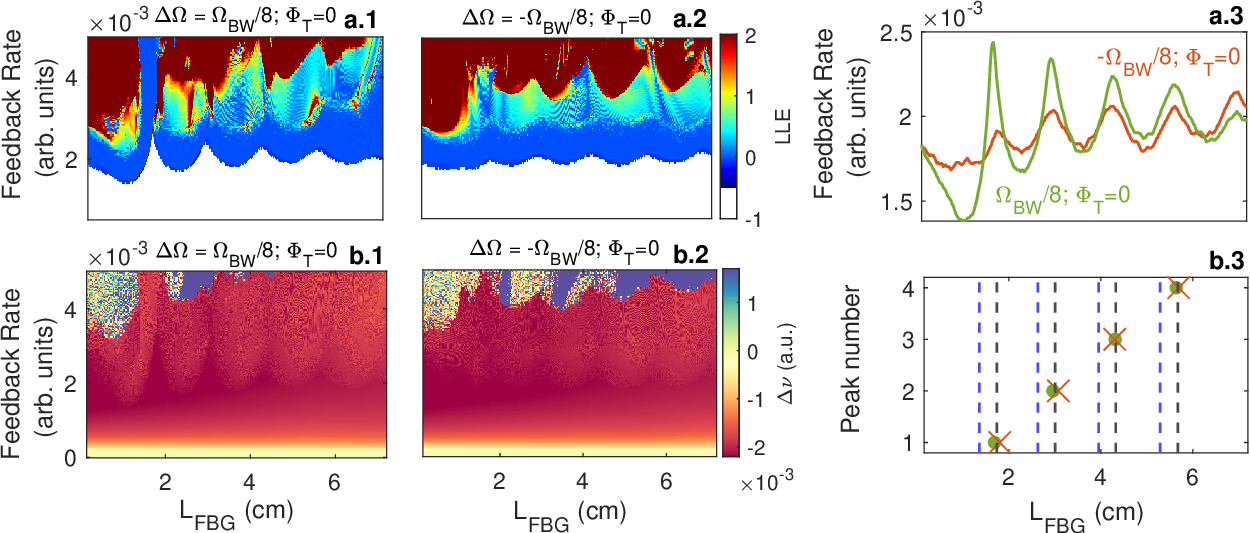}
\end{center}
\caption{\label{fig:FBGphase} Numerical map of the values of the LLE (a.1.) as a function of $L_\textrm{FBG}$ and feedback rate with $\Phi_\textrm{T}=0$ for detuning $\Omega_{BW}/8$ (a.1), and -$\Omega_{BW}/8$ (a.2). Variations of the emission frequency of the laser $\Delta\nu$ for detuning $\Omega_{BW}/8$ (b.1), and -$\Omega_{BW}/8$ (b.2). The evolution of the Hopf bifurcation with $L_\textrm{FBG}$ for both detunings at feedback phase $\Phi_\textrm{T}=0$ (a.3) and the corresponding stability peaks extracted from the simulation (green dots-positive detuning; orange crosses-negative detuning) and calculated from the theory (blue dashed lines-closer peak; black dashed lines-further peak).}
\end{figure*}

\begin{figure}
\begin{center}
\includegraphics[width=\linewidth]{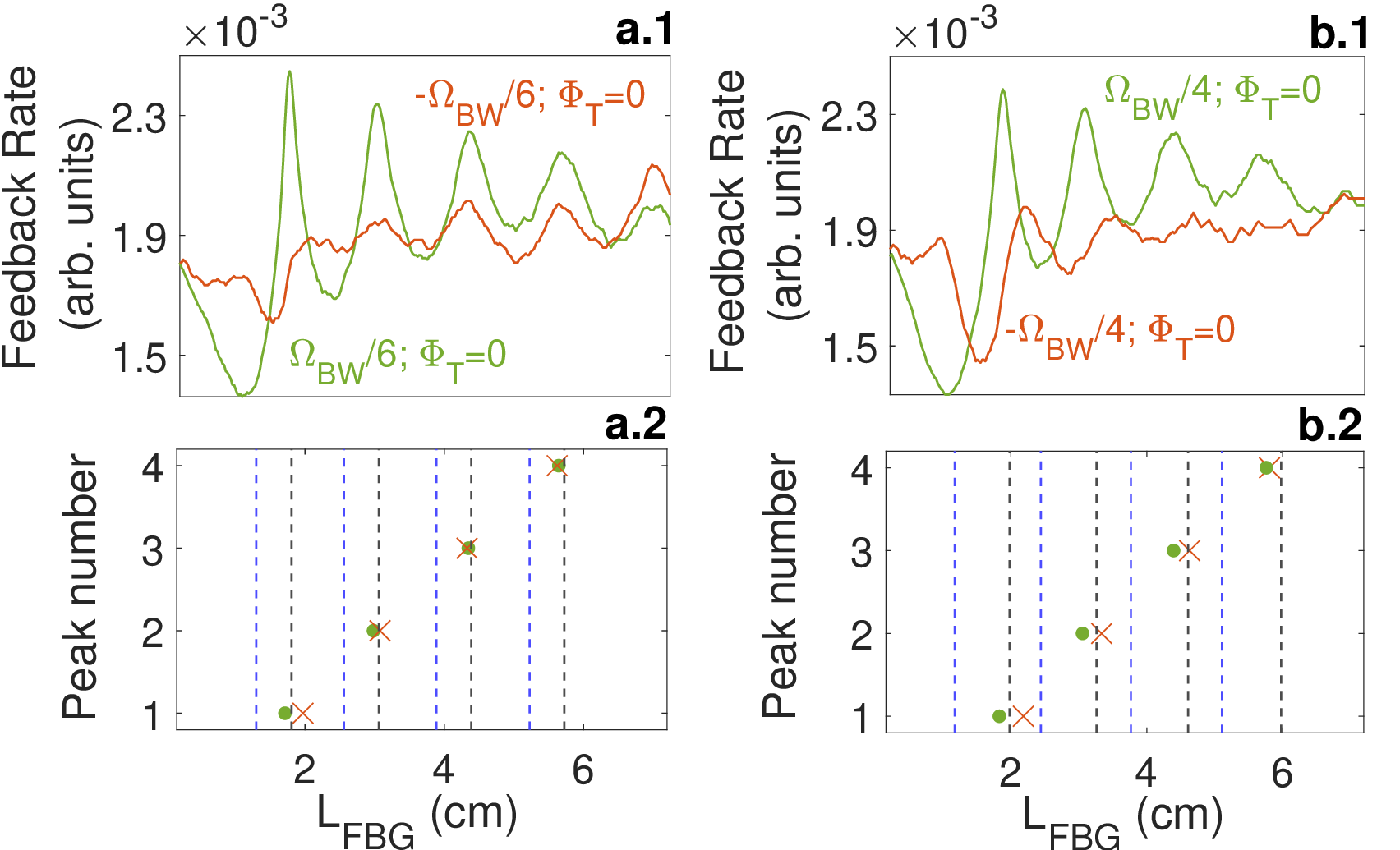}
\end{center}
\caption{\label{fig:FBGphase2} The evolution of the Hopf bifurcation with $L_\textrm{FBG}$ $\Phi_\textrm{T}=0$ for detuning $\pm\Omega_{BW}/6$ (a.1), $\pm\Omega_{BW}/4$ (b.1) and the corresponding stability peaks extracted from the simulation (green dots-positive detuning; orange crosses-negative detunings) and calculated from the theory (blue dashed lines-closer peak; black dashed lines-further peak) for $\pm\Omega_{BW}/4$ (a.2), $\pm\Omega_{BW}/4$ (b.2).}
\end{figure}

In contrast to conventional filtered optical feedback, FBG feedback involves an intrinsic reflection phase that affects the total feedback phase and subsequently the dynamics of the laser. This FBG reflection phase, given as $\Phi_\textrm{FBG}=\textrm{arg}\left(r(\Omega)/r(0)\right)$, is depicted in orange in FIG. \ref{fig:FBGphaseSchematics}. The shape of the $\Phi_\textrm{FBG}$ component changes for positive and negative detuning and exhibits rapid variations near the edges of the lobes due to chromatic dispersion, which has been reported in various grating structures \cite{618322}. As the FBG phase has a distinct shape and sign for positive and negative detuning, it is likely that the laser will exhibit different behavior for these two cases, leading to an asymmetry of the dynamics with respect to the center frequency. To analyze the impact of the $\Phi_\textrm{FBG}$ component, we consider detunings of $\Omega_{BW}/8$ and -$\Omega_{BW}/8$ and vary the offset phase of the system in each case such that the total phase of the system is zero. The $\Phi_\textrm{FBG}$ component can be estimated using the following equation:
\begin{equation}
  \Phi_\textrm{T}= \Phi_\textrm{FBG} + \Phi_0
\label{eq:fbgPhase}  
\end{equation}
where $\Phi_\textrm{T}$ is the total phase of the system, and $\Phi_0$ is the offset phase. Thus, to compensate the effects of $\Phi_\textrm{FBG}$ we vary the offset phase to counter-act the intrinsic phase of the FBG. 
The overall dynamics of the laser are again analyzed by the LLE and the numerical maps corresponding to the positive and negative detuning are shown in FIG. \ref{fig:FBGphase}(a.1) and (a.2) respectively. The evolution of the Hopf bifurcation for both cases is shown in FIG. \ref{fig:FBGphase}(a.3). Notably, we observe that the position of the stabilization peaks is now the same for both detunings. This observation holds true for detunings of $\Omega_{BW}/6$ and -$\Omega_{BW}/6$ as shown in FIG. \ref{fig:FBGphase2}(a.1) and (a.2). Furthermore, the values obtained for detunings of $\Omega_{BW}/4$ and -$\Omega_{BW}/4$ are much closer to the theoretically calculated further peak, as depicted in FIG. \ref{fig:FBGphase2}(b.1) and (b.2), although with a remaining yet smaller deviation
These findings suggest that the asymmetry between positive and negative detunings is primarily influenced by the phase characteristics of the FBG. In other words, the fluctuations in stability in the case of FBG feedback are not solely determined by the damping of the laser's relaxation oscillations but are also regulated by the intrinsic phase of the FBG. 
However, the remaining deviation observed for detuning $\Omega_{BW}/4$ and -$\Omega_{BW}/4$ indicates the influence of other factors on the asymmetry between positive and negative detunings, apart from the FBG phase. Additionally, there are differences in the height and sharpness of the peaks for positive and negative detunings. Moreover, from FIG. \ref{fig:FBGphase}(a.1) and (a.2), we observe that although the stability peak positions coincide, the most complex dynamics differ.

\subsection{Shift of the laser frequency}
\label{sec:delta_nu}
In previous studies, it has been demonstrated that optical feedback power into the laser causes a reduction in optical gain, leading to a shift in the laser's emission frequency \cite{5419246, 4633714}. This phenomenon may have implications for the laser's behavior and could contribute to the observed asymmetry between positive and negative detunings. To explore this in more detail, we calculate the frequency shift by analyzing the variation of the optical field phase.
The numerical mapping of the emission frequency variation, denoted as $\Delta\nu$, is presented as a function of $L_\textrm{FBG}$ and feedback rate in FIG. \ref{fig:FBGphase}(b.1) and (b.2) for frequency detunings of $\Omega_{BW}/8$ and -$\Omega_{BW}/8$ respectively. It is worth noting that the variation of the emission frequency follows a similar trend as the variations observed in the LLE under the same conditions, but it differs between positive and negative detunings.
The variation in emission frequency will impact the position of the stability peaks because the side lobes of the relaxation oscillations, and thus their overlap with the zeros of the FBG reflectivity spectrum, will shift by an amount $\Delta\nu$. Additionally, since the variation of emission frequency differs between positive and negative detunings, as shown in FIG. \ref{fig:FBGphase}(b.1) and (b.2), the occurrence of stability peaks will also differ and exhibit asymmetry with respect to the central frequency of the filter.

To assess the impact of the frequency shift $\Delta\nu$, for calculation purposes we express Eq. \ref{eq:L_detuned} as:
\begin{equation}
  L_\textrm{FBG}^{(n)}=\frac{\pi c \left(\sqrt{n^2+\left(\frac{\kappa L_\textrm{FBG}}{\pi}\right)^2}\right)}{n_\textrm{eff}\left(\Omega_{RO}\pm\Delta\Omega - 2\pi\Delta\nu \right)}
\label{eq:L_detuned_plus_delta_nu}  
\end{equation}
Although the frequency shift $\Delta\nu$ exhibits a seemingly small variation, which differs for positive and negative detunings, it alone cannot fully explain the disparity observed between the simulated and theoretically calculated stability peaks (see FIG. \ref{fig:FBGphase}(b.1) and (b.2)). The influence of this minor variation on the position of the stability peaks is minimal, to the extent that creating a plot similar to FIG. \ref{fig:FBGphase2}(b.1) and (b.2) would render the change barely visible. Thus, while the frequency shift plays a role, other contributing factors must account for the asymmetry in the dynamics resulting from the frequency detuning.

\section{Conclusion} 
\label{sec:conclusion} 
In this paper, we have explored the effect of feedback time-distribution on laser dynamics using FBG feedback as testbed. We found that the distribution of the feedback in time has limited impact on the laser stability for gratings shorter than $1$\,cm. Yet, for longer gratings, fluctuations in the laser stability emerged. Like in the case of filtered optical feedback, this behavior could be explained in part by considering the alignment of the zeros of the FBG reflectivity spectrum and the laser side lobes around the relaxation oscillation frequency. Similarly, these stability fluctuations are present when the frequency detuning between the Bragg wavelength of the FBG and the free running frequency of the laser is changed, however, they are not symmetric with respect to the Bragg wavelength. To better understand this asymmetry, we analyzed the feedback parameters such as the FBG phase, the feedback-induced frequency shift, along with the variations of the relaxation oscillation frequency. We have demonstrated that the position of the stability peaks coincides for both positive and negative detunings when compensating for the impact of the FBG phase. Our findings essentially show that the stability of lasers with FBG feedback shares similarities with the case of filtered optical feedback, but that the precise shape of the FBG spectrum and, more importantly, the phase of the FBG add to the complex dynamics. The asymmetrical behavior with respect to the detuning seems to be an intrinsic characteristic of the FBG feedback.  

It is important to note that at high feedback rates, the dynamics become more intricate, and the conclusions drawn here may not hold true. Experimental verification is necessary to validate our results. Nevertheless, conducting such experiments presents considerable challenges, particularly when it comes to implementing the synthetic control of the FBG phase employed in our analysis.  

In conclusion, we have provided insights into the asymmetrical behavior observed in the presence of FBG feedback and identified the influence of the FBG phase, relaxation oscillations, and offset phase on stability regions and feedback rates. These findings contribute to a better understanding of the complex behavior exhibited by lasers with FBG feedback and provide valuable guidance for further investigations in this field. 

\begin{acknowledgments}
We acknowledge the support of Fonds Wetenschappelijk Onderzoek (FIONA, project number G029619N, COLOR’UP, project number G0G0319N) and the METHUSALEM program of the Flemish Government (Vlaamse Overheid).

We would like to acknowledge the assistance provided by Ir. Mennatallah A. Z. Kandil at the beginnings of this work as part of her master thesis project.   
\end{acknowledgments}

\bibliography{report}

\begin{thebibliography}{45}%
\makeatletter
\providecommand \@ifxundefined [1]{%
 \@ifx{#1\undefined}
}%
\providecommand \@ifnum [1]{%
 \ifnum #1\expandafter \@firstoftwo
 \else \expandafter \@secondoftwo
 \fi
}%
\providecommand \@ifx [1]{%
 \ifx #1\expandafter \@firstoftwo
 \else \expandafter \@secondoftwo
 \fi
}%
\providecommand \natexlab [1]{#1}%
\providecommand \enquote  [1]{``#1''}%
\providecommand \bibnamefont  [1]{#1}%
\providecommand \bibfnamefont [1]{#1}%
\providecommand \citenamefont [1]{#1}%
\providecommand \href@noop [0]{\@secondoftwo}%
\providecommand \href [0]{\begingroup \@sanitize@url \@href}%
\providecommand \@href[1]{\@@startlink{#1}\@@href}%
\providecommand \@@href[1]{\endgroup#1\@@endlink}%
\providecommand \@sanitize@url [0]{\catcode `\\12\catcode `\$12\catcode
  `\&12\catcode `\#12\catcode `\^12\catcode `\_12\catcode `\%12\relax}%
\providecommand \@@startlink[1]{}%
\providecommand \@@endlink[0]{}%
\providecommand \url  [0]{\begingroup\@sanitize@url \@url }%
\providecommand \@url [1]{\endgroup\@href {#1}{\urlprefix }}%
\providecommand \urlprefix  [0]{URL }%
\providecommand \Eprint [0]{\href }%
\providecommand \doibase [0]{https://doi.org/}%
\providecommand \selectlanguage [0]{\@gobble}%
\providecommand \bibinfo  [0]{\@secondoftwo}%
\providecommand \bibfield  [0]{\@secondoftwo}%
\providecommand \translation [1]{[#1]}%
\providecommand \BibitemOpen [0]{}%
\providecommand \bibitemStop [0]{}%
\providecommand \bibitemNoStop [0]{.\EOS\space}%
\providecommand \EOS [0]{\spacefactor3000\relax}%
\providecommand \BibitemShut  [1]{\csname bibitem#1\endcsname}%
\let\auto@bib@innerbib\@empty
%</preamble>
\bibitem [{\citenamefont {Soriano}\ \emph {et~al.}(2013)\citenamefont
  {Soriano}, \citenamefont {Garc\'{\i}a-Ojalvo}, \citenamefont {Mirasso},\ and\
  \citenamefont {Fischer}}]{RevModPhys.85.421}%
  \BibitemOpen
  \bibfield  {author} {\bibinfo {author} {\bibfnamefont {M.~C.}\ \bibnamefont
  {Soriano}}, \bibinfo {author} {\bibfnamefont {J.}~\bibnamefont
  {Garc\'{\i}a-Ojalvo}}, \bibinfo {author} {\bibfnamefont {C.~R.}\ \bibnamefont
  {Mirasso}},\ and\ \bibinfo {author} {\bibfnamefont {I.}~\bibnamefont
  {Fischer}},\ }\bibfield  {title} {\bibinfo {title} {Complex photonics:
  Dynamics and applications of delay-coupled semiconductors lasers},\ }\href
  {https://doi.org/10.1103/RevModPhys.85.421} {\bibfield  {journal} {\bibinfo
  {journal} {Rev. Mod. Phys.}\ }\textbf {\bibinfo {volume} {85}},\ \bibinfo
  {pages} {421} (\bibinfo {year} {2013})}\BibitemShut {NoStop}%
\bibitem [{\citenamefont {Wu}\ \emph {et~al.}(2013)\citenamefont {Wu},
  \citenamefont {Wu}, \citenamefont {Liu}, \citenamefont {Fan}, \citenamefont
  {Tang},\ and\ \citenamefont {Xia}}]{Wu:13}%
  \BibitemOpen
  \bibfield  {author} {\bibinfo {author} {\bibfnamefont {J.-G.}\ \bibnamefont
  {Wu}}, \bibinfo {author} {\bibfnamefont {Z.-M.}\ \bibnamefont {Wu}}, \bibinfo
  {author} {\bibfnamefont {Y.-R.}\ \bibnamefont {Liu}}, \bibinfo {author}
  {\bibfnamefont {L.}~\bibnamefont {Fan}}, \bibinfo {author} {\bibfnamefont
  {X.}~\bibnamefont {Tang}},\ and\ \bibinfo {author} {\bibfnamefont {G.-Q.}\
  \bibnamefont {Xia}},\ }\bibfield  {title} {\bibinfo {title} {Simulation of
  bidirectional long-distance chaos communication performance in a novel
  fiber-optic chaos synchronization system},\ }\href
  {http://opg.optica.org/jlt/abstract.cfm?URI=jlt-31-3-461} {\bibfield
  {journal} {\bibinfo  {journal} {J. Lightwave Technol.}\ }\textbf {\bibinfo
  {volume} {31}},\ \bibinfo {pages} {461} (\bibinfo {year} {2013})}\BibitemShut
  {NoStop}%
\bibitem [{\citenamefont {Xue}\ \emph {et~al.}(2016)\citenamefont {Xue},
  \citenamefont {Jiang}, \citenamefont {Lv}, \citenamefont {Wang},
  \citenamefont {Li}, \citenamefont {Lin},\ and\ \citenamefont {Qiu}}]{Xue:16}%
  \BibitemOpen
  \bibfield  {author} {\bibinfo {author} {\bibfnamefont {C.}~\bibnamefont
  {Xue}}, \bibinfo {author} {\bibfnamefont {N.}~\bibnamefont {Jiang}}, \bibinfo
  {author} {\bibfnamefont {Y.}~\bibnamefont {Lv}}, \bibinfo {author}
  {\bibfnamefont {C.}~\bibnamefont {Wang}}, \bibinfo {author} {\bibfnamefont
  {G.}~\bibnamefont {Li}}, \bibinfo {author} {\bibfnamefont {S.}~\bibnamefont
  {Lin}},\ and\ \bibinfo {author} {\bibfnamefont {K.}~\bibnamefont {Qiu}},\
  }\bibfield  {title} {\bibinfo {title} {Security-enhanced chaos communication
  with time-delay signature suppression and phase encryption},\ }\href
  {https://doi.org/10.1364/OL.41.003690} {\bibfield  {journal} {\bibinfo
  {journal} {Opt. Lett.}\ }\textbf {\bibinfo {volume} {41}},\ \bibinfo {pages}
  {3690} (\bibinfo {year} {2016})}\BibitemShut {NoStop}%
\bibitem [{\citenamefont {Li}\ \emph {et~al.}(2017)\citenamefont {Li},
  \citenamefont {Zhang}, \citenamefont {Sang}, \citenamefont {Liu},
  \citenamefont {Guo}, \citenamefont {Guo}, \citenamefont {Wang}, \citenamefont
  {Shore},\ and\ \citenamefont {Wang}}]{Li:17}%
  \BibitemOpen
  \bibfield  {author} {\bibinfo {author} {\bibfnamefont {P.}~\bibnamefont
  {Li}}, \bibinfo {author} {\bibfnamefont {J.}~\bibnamefont {Zhang}}, \bibinfo
  {author} {\bibfnamefont {L.}~\bibnamefont {Sang}}, \bibinfo {author}
  {\bibfnamefont {X.}~\bibnamefont {Liu}}, \bibinfo {author} {\bibfnamefont
  {Y.}~\bibnamefont {Guo}}, \bibinfo {author} {\bibfnamefont {X.}~\bibnamefont
  {Guo}}, \bibinfo {author} {\bibfnamefont {A.}~\bibnamefont {Wang}}, \bibinfo
  {author} {\bibfnamefont {K.~A.}\ \bibnamefont {Shore}},\ and\ \bibinfo
  {author} {\bibfnamefont {Y.}~\bibnamefont {Wang}},\ }\bibfield  {title}
  {\bibinfo {title} {Real-time online photonic random number generation},\
  }\href {https://doi.org/10.1364/OL.42.002699} {\bibfield  {journal} {\bibinfo
   {journal} {Opt. Lett.}\ }\textbf {\bibinfo {volume} {42}},\ \bibinfo {pages}
  {2699} (\bibinfo {year} {2017})}\BibitemShut {NoStop}%
\bibitem [{\citenamefont {Argyris}\ \emph {et~al.}(2016)\citenamefont
  {Argyris}, \citenamefont {Pikasis},\ and\ \citenamefont
  {Syvridis}}]{Argyris:16}%
  \BibitemOpen
  \bibfield  {author} {\bibinfo {author} {\bibfnamefont {A.}~\bibnamefont
  {Argyris}}, \bibinfo {author} {\bibfnamefont {E.}~\bibnamefont {Pikasis}},\
  and\ \bibinfo {author} {\bibfnamefont {D.}~\bibnamefont {Syvridis}},\
  }\bibfield  {title} {\bibinfo {title} {Gb/s one-time-pad data encryption with
  synchronized chaos-based true random bit generators},\ }\href
  {http://opg.optica.org/jlt/abstract.cfm?URI=jlt-34-22-5325} {\bibfield
  {journal} {\bibinfo  {journal} {J. Lightwave Technol.}\ }\textbf {\bibinfo
  {volume} {34}},\ \bibinfo {pages} {5325} (\bibinfo {year}
  {2016})}\BibitemShut {NoStop}%
\bibitem [{\citenamefont {Li}\ \emph {et~al.}(2016)\citenamefont {Li},
  \citenamefont {Sun}, \citenamefont {Liu}, \citenamefont {Yi}, \citenamefont
  {Zhang}, \citenamefont {Guo}, \citenamefont {Guo},\ and\ \citenamefont
  {Wang}}]{Li:16}%
  \BibitemOpen
  \bibfield  {author} {\bibinfo {author} {\bibfnamefont {P.}~\bibnamefont
  {Li}}, \bibinfo {author} {\bibfnamefont {Y.}~\bibnamefont {Sun}}, \bibinfo
  {author} {\bibfnamefont {X.}~\bibnamefont {Liu}}, \bibinfo {author}
  {\bibfnamefont {X.}~\bibnamefont {Yi}}, \bibinfo {author} {\bibfnamefont
  {J.}~\bibnamefont {Zhang}}, \bibinfo {author} {\bibfnamefont
  {X.}~\bibnamefont {Guo}}, \bibinfo {author} {\bibfnamefont {Y.}~\bibnamefont
  {Guo}},\ and\ \bibinfo {author} {\bibfnamefont {Y.}~\bibnamefont {Wang}},\
  }\bibfield  {title} {\bibinfo {title} {Fully photonics-based physical random
  bit generator},\ }\href {https://doi.org/10.1364/OL.41.003347} {\bibfield
  {journal} {\bibinfo  {journal} {Opt. Lett.}\ }\textbf {\bibinfo {volume}
  {41}},\ \bibinfo {pages} {3347} (\bibinfo {year} {2016})}\BibitemShut
  {NoStop}%
\bibitem [{\citenamefont {Ugajin}\ \emph {et~al.}(2017)\citenamefont {Ugajin},
  \citenamefont {Terashima}, \citenamefont {Iwakawa}, \citenamefont {Uchida},
  \citenamefont {Harayama}, \citenamefont {Yoshimura},\ and\ \citenamefont
  {Inubushi}}]{Ugajin:17}%
  \BibitemOpen
  \bibfield  {author} {\bibinfo {author} {\bibfnamefont {K.}~\bibnamefont
  {Ugajin}}, \bibinfo {author} {\bibfnamefont {Y.}~\bibnamefont {Terashima}},
  \bibinfo {author} {\bibfnamefont {K.}~\bibnamefont {Iwakawa}}, \bibinfo
  {author} {\bibfnamefont {A.}~\bibnamefont {Uchida}}, \bibinfo {author}
  {\bibfnamefont {T.}~\bibnamefont {Harayama}}, \bibinfo {author}
  {\bibfnamefont {K.}~\bibnamefont {Yoshimura}},\ and\ \bibinfo {author}
  {\bibfnamefont {M.}~\bibnamefont {Inubushi}},\ }\bibfield  {title} {\bibinfo
  {title} {Real-time fast physical random number generator with a photonic
  integrated circuit},\ }\href {https://doi.org/10.1364/OE.25.006511}
  {\bibfield  {journal} {\bibinfo  {journal} {Opt. Express}\ }\textbf {\bibinfo
  {volume} {25}},\ \bibinfo {pages} {6511} (\bibinfo {year}
  {2017})}\BibitemShut {NoStop}%
\bibitem [{\citenamefont {Lin}\ and\ \citenamefont {Liu}(2004)}]{1303798}%
  \BibitemOpen
  \bibfield  {author} {\bibinfo {author} {\bibfnamefont {F.-Y.}\ \bibnamefont
  {Lin}}\ and\ \bibinfo {author} {\bibfnamefont {J.-M.}\ \bibnamefont {Liu}},\
  }\bibfield  {title} {\bibinfo {title} {Chaotic radar using nonlinear laser
  dynamics},\ }\href {https://doi.org/10.1109/JQE.2004.828237} {\bibfield
  {journal} {\bibinfo  {journal} {IEEE Journal of Quantum Electronics}\
  }\textbf {\bibinfo {volume} {40}},\ \bibinfo {pages} {815} (\bibinfo {year}
  {2004})}\BibitemShut {NoStop}%
\bibitem [{\citenamefont {Xiang}\ \emph {et~al.}(2019)\citenamefont {Xiang},
  \citenamefont {Zhang}, \citenamefont {Gong}, \citenamefont {Guo},
  \citenamefont {Lin},\ and\ \citenamefont {Hao}}]{8693533}%
  \BibitemOpen
  \bibfield  {author} {\bibinfo {author} {\bibfnamefont {S.}~\bibnamefont
  {Xiang}}, \bibinfo {author} {\bibfnamefont {Y.}~\bibnamefont {Zhang}},
  \bibinfo {author} {\bibfnamefont {J.}~\bibnamefont {Gong}}, \bibinfo {author}
  {\bibfnamefont {X.}~\bibnamefont {Guo}}, \bibinfo {author} {\bibfnamefont
  {L.}~\bibnamefont {Lin}},\ and\ \bibinfo {author} {\bibfnamefont
  {Y.}~\bibnamefont {Hao}},\ }\bibfield  {title} {\bibinfo {title} {Stdp-based
  unsupervised spike pattern learning in a photonic spiking neural network with
  vcsels and vcsoas},\ }\href {https://doi.org/10.1109/JSTQE.2019.2911565}
  {\bibfield  {journal} {\bibinfo  {journal} {IEEE Journal of Selected Topics
  in Quantum Electronics}\ }\textbf {\bibinfo {volume} {25}},\ \bibinfo {pages}
  {1} (\bibinfo {year} {2019})}\BibitemShut {NoStop}%
\bibitem [{\citenamefont {Oliver}\ \emph {et~al.}(2011)\citenamefont {Oliver},
  \citenamefont {Soriano}, \citenamefont {Sukow},\ and\ \citenamefont
  {Fischer}}]{Oliver:11}%
  \BibitemOpen
  \bibfield  {author} {\bibinfo {author} {\bibfnamefont {N.}~\bibnamefont
  {Oliver}}, \bibinfo {author} {\bibfnamefont {M.~C.}\ \bibnamefont {Soriano}},
  \bibinfo {author} {\bibfnamefont {D.~W.}\ \bibnamefont {Sukow}},\ and\
  \bibinfo {author} {\bibfnamefont {I.}~\bibnamefont {Fischer}},\ }\bibfield
  {title} {\bibinfo {title} {Dynamics of a semiconductor laser with
  polarization-rotated feedback and its utilization for random bit
  generation},\ }\href {https://doi.org/10.1364/OL.36.004632} {\bibfield
  {journal} {\bibinfo  {journal} {Opt. Lett.}\ }\textbf {\bibinfo {volume}
  {36}},\ \bibinfo {pages} {4632} (\bibinfo {year} {2011})}\BibitemShut
  {NoStop}%
\bibitem [{\citenamefont {Friart}\ \emph {et~al.}(2014)\citenamefont {Friart},
  \citenamefont {Weicker}, \citenamefont {Danckaert},\ and\ \citenamefont
  {Erneux}}]{Friart:14}%
  \BibitemOpen
  \bibfield  {author} {\bibinfo {author} {\bibfnamefont {G.}~\bibnamefont
  {Friart}}, \bibinfo {author} {\bibfnamefont {L.}~\bibnamefont {Weicker}},
  \bibinfo {author} {\bibfnamefont {J.}~\bibnamefont {Danckaert}},\ and\
  \bibinfo {author} {\bibfnamefont {T.}~\bibnamefont {Erneux}},\ }\bibfield
  {title} {\bibinfo {title} {Relaxation and square-wave oscillations in a
  semiconductor laser with polarization rotated optical feedback},\ }\href
  {https://doi.org/10.1364/OE.22.006905} {\bibfield  {journal} {\bibinfo
  {journal} {Opt. Express}\ }\textbf {\bibinfo {volume} {22}},\ \bibinfo
  {pages} {6905} (\bibinfo {year} {2014})}\BibitemShut {NoStop}%
\bibitem [{\citenamefont {Lawrence}\ and\ \citenamefont
  {Kane}(2001)}]{PhysRevA.63.033805}%
  \BibitemOpen
  \bibfield  {author} {\bibinfo {author} {\bibfnamefont {J.~S.}\ \bibnamefont
  {Lawrence}}\ and\ \bibinfo {author} {\bibfnamefont {D.~M.}\ \bibnamefont
  {Kane}},\ }\bibfield  {title} {\bibinfo {title} {Contrasting conventional
  optical and phase-conjugate feedback in laser diodes},\ }\href
  {https://doi.org/10.1103/PhysRevA.63.033805} {\bibfield  {journal} {\bibinfo
  {journal} {Phys. Rev. A}\ }\textbf {\bibinfo {volume} {63}},\ \bibinfo
  {pages} {033805} (\bibinfo {year} {2001})}\BibitemShut {NoStop}%
\bibitem [{\citenamefont {Virte}\ \emph {et~al.}(2011)\citenamefont {Virte},
  \citenamefont {Karsaklian Dal~Bosco}, \citenamefont {Wolfersberger},\ and\
  \citenamefont {Sciamanna}}]{PhysRevA.84.043836}%
  \BibitemOpen
  \bibfield  {author} {\bibinfo {author} {\bibfnamefont {M.}~\bibnamefont
  {Virte}}, \bibinfo {author} {\bibfnamefont {A.}~\bibnamefont {Karsaklian
  Dal~Bosco}}, \bibinfo {author} {\bibfnamefont {D.}~\bibnamefont
  {Wolfersberger}},\ and\ \bibinfo {author} {\bibfnamefont {M.}~\bibnamefont
  {Sciamanna}},\ }\bibfield  {title} {\bibinfo {title} {Chaos crisis and
  bistability of self-pulsing dynamics in a laser diode with phase-conjugate
  feedback},\ }\href {https://doi.org/10.1103/PhysRevA.84.043836} {\bibfield
  {journal} {\bibinfo  {journal} {Phys. Rev. A}\ }\textbf {\bibinfo {volume}
  {84}},\ \bibinfo {pages} {043836} (\bibinfo {year} {2011})}\BibitemShut
  {NoStop}%
\bibitem [{\citenamefont {Malica}\ \emph {et~al.}(2020)\citenamefont {Malica},
  \citenamefont {Bouchez}, \citenamefont {Wolfersberger},\ and\ \citenamefont
  {Sciamanna}}]{Malica:20}%
  \BibitemOpen
  \bibfield  {author} {\bibinfo {author} {\bibfnamefont {T.}~\bibnamefont
  {Malica}}, \bibinfo {author} {\bibfnamefont {G.}~\bibnamefont {Bouchez}},
  \bibinfo {author} {\bibfnamefont {D.}~\bibnamefont {Wolfersberger}},\ and\
  \bibinfo {author} {\bibfnamefont {M.}~\bibnamefont {Sciamanna}},\ }\bibfield
  {title} {\bibinfo {title} {Spatiotemporal complexity of chaos in a
  phase-conjugate feedback laser system},\ }\href
  {https://doi.org/10.1364/OL.383557} {\bibfield  {journal} {\bibinfo
  {journal} {Opt. Lett.}\ }\textbf {\bibinfo {volume} {45}},\ \bibinfo {pages}
  {819} (\bibinfo {year} {2020})}\BibitemShut {NoStop}%
\bibitem [{\citenamefont {Ma}\ \emph {et~al.}(2020)\citenamefont {Ma},
  \citenamefont {Xiang}, \citenamefont {Guo}, \citenamefont {Song},
  \citenamefont {Wen},\ and\ \citenamefont {Hao}}]{Ma:20}%
  \BibitemOpen
  \bibfield  {author} {\bibinfo {author} {\bibfnamefont {Y.}~\bibnamefont
  {Ma}}, \bibinfo {author} {\bibfnamefont {S.}~\bibnamefont {Xiang}}, \bibinfo
  {author} {\bibfnamefont {X.}~\bibnamefont {Guo}}, \bibinfo {author}
  {\bibfnamefont {Z.}~\bibnamefont {Song}}, \bibinfo {author} {\bibfnamefont
  {A.}~\bibnamefont {Wen}},\ and\ \bibinfo {author} {\bibfnamefont
  {Y.}~\bibnamefont {Hao}},\ }\bibfield  {title} {\bibinfo {title} {Time-delay
  signature concealment of chaos and ultrafast decision making in mutually
  coupled semiconductor lasers with a phase-modulated sagnac loop},\ }\href
  {https://doi.org/10.1364/OE.384378} {\bibfield  {journal} {\bibinfo
  {journal} {Opt. Express}\ }\textbf {\bibinfo {volume} {28}},\ \bibinfo
  {pages} {1665} (\bibinfo {year} {2020})}\BibitemShut {NoStop}%
\bibitem [{\citenamefont {Kovalev}\ \emph {et~al.}(2021)\citenamefont
  {Kovalev}, \citenamefont {Viktorov},\ and\ \citenamefont
  {Erneux}}]{PhysRevE.103.042206}%
  \BibitemOpen
  \bibfield  {author} {\bibinfo {author} {\bibfnamefont {A.~V.}\ \bibnamefont
  {Kovalev}}, \bibinfo {author} {\bibfnamefont {E.~A.}\ \bibnamefont
  {Viktorov}},\ and\ \bibinfo {author} {\bibfnamefont {T.}~\bibnamefont
  {Erneux}},\ }\bibfield  {title} {\bibinfo {title} {Short-delayed-feedback
  semiconductor lasers},\ }\href {https://doi.org/10.1103/PhysRevE.103.042206}
  {\bibfield  {journal} {\bibinfo  {journal} {Phys. Rev. E}\ }\textbf {\bibinfo
  {volume} {103}},\ \bibinfo {pages} {042206} (\bibinfo {year}
  {2021})}\BibitemShut {NoStop}%
\bibitem [{\citenamefont {Li}\ \emph {et~al.}(2012)\citenamefont {Li},
  \citenamefont {Liu},\ and\ \citenamefont {Chan}}]{6320711}%
  \BibitemOpen
  \bibfield  {author} {\bibinfo {author} {\bibfnamefont {S.-S.}\ \bibnamefont
  {Li}}, \bibinfo {author} {\bibfnamefont {Q.}~\bibnamefont {Liu}},\ and\
  \bibinfo {author} {\bibfnamefont {S.-C.}\ \bibnamefont {Chan}},\ }\bibfield
  {title} {\bibinfo {title} {Distributed feedbacks for time-delay signature
  suppression of chaos generated from a semiconductor laser},\ }\href
  {https://doi.org/10.1109/JPHOT.2012.2220759} {\bibfield  {journal} {\bibinfo
  {journal} {IEEE Photonics Journal}\ }\textbf {\bibinfo {volume} {4}},\
  \bibinfo {pages} {1930} (\bibinfo {year} {2012})}\BibitemShut {NoStop}%
\bibitem [{\citenamefont {Li}\ and\ \citenamefont {Chan}(2015)}]{7097638}%
  \BibitemOpen
  \bibfield  {author} {\bibinfo {author} {\bibfnamefont {S.-S.}\ \bibnamefont
  {Li}}\ and\ \bibinfo {author} {\bibfnamefont {S.-C.}\ \bibnamefont {Chan}},\
  }\bibfield  {title} {\bibinfo {title} {Chaotic time-delay signature
  suppression in a semiconductor laser with frequency-detuned grating
  feedback},\ }\href {https://doi.org/10.1109/JSTQE.2015.2427521} {\bibfield
  {journal} {\bibinfo  {journal} {IEEE Journal of Selected Topics in Quantum
  Electronics}\ }\textbf {\bibinfo {volume} {21}},\ \bibinfo {pages} {541}
  (\bibinfo {year} {2015})}\BibitemShut {NoStop}%
\bibitem [{\citenamefont {Jun-Feng}\ \emph {et~al.}(2017)\citenamefont
  {Jun-Feng}, \citenamefont {Zhu-Qiang}, \citenamefont {Guang-Na},
  \citenamefont {Guang-Qiong},\ and\ \citenamefont
  {Zheng-Mao}}]{jun-feng_characteristics_2017}%
  \BibitemOpen
  \bibfield  {author} {\bibinfo {author} {\bibfnamefont {Q.}~\bibnamefont
  {Jun-Feng}}, \bibinfo {author} {\bibfnamefont {Z.}~\bibnamefont {Zhu-Qiang}},
  \bibinfo {author} {\bibfnamefont {W.}~\bibnamefont {Guang-Na}}, \bibinfo
  {author} {\bibfnamefont {X.}~\bibnamefont {Guang-Qiong}},\ and\ \bibinfo
  {author} {\bibfnamefont {W.}~\bibnamefont {Zheng-Mao}},\ }\bibfield  {title}
  {\bibinfo {title} {Characteristics of chaotic output from a {Gaussian}
  apodized fiber {Bragg} grating external-cavity semiconductor laser},\
  }\href@noop {} {\bibfield  {journal} {\bibinfo  {journal} {Acta physica
  Sinica}\ }\textbf {\bibinfo {volume} {66}},\ \bibinfo {pages} {244207}
  (\bibinfo {year} {2017})}\BibitemShut {NoStop}%
\bibitem [{\citenamefont {Wang}\ \emph {et~al.}(2017)\citenamefont {Wang},
  \citenamefont {Wang}, \citenamefont {Li},\ and\ \citenamefont
  {Wang}}]{Wang:17}%
  \BibitemOpen
  \bibfield  {author} {\bibinfo {author} {\bibfnamefont {A.}~\bibnamefont
  {Wang}}, \bibinfo {author} {\bibfnamefont {L.}~\bibnamefont {Wang}}, \bibinfo
  {author} {\bibfnamefont {P.}~\bibnamefont {Li}},\ and\ \bibinfo {author}
  {\bibfnamefont {Y.}~\bibnamefont {Wang}},\ }\bibfield  {title} {\bibinfo
  {title} {Minimal-post-processing 320-gbps true random bit generation using
  physical white chaos},\ }\href {https://doi.org/10.1364/OE.25.003153}
  {\bibfield  {journal} {\bibinfo  {journal} {Opt. Express}\ }\textbf {\bibinfo
  {volume} {25}},\ \bibinfo {pages} {3153} (\bibinfo {year}
  {2017})}\BibitemShut {NoStop}%
\bibitem [{\citenamefont {Wang}\ \emph {et~al.}(2019)\citenamefont {Wang},
  \citenamefont {Wang}, \citenamefont {Guo}, \citenamefont {Wang},\ and\
  \citenamefont {Wang}}]{Wang:19}%
  \BibitemOpen
  \bibfield  {author} {\bibinfo {author} {\bibfnamefont {D.~M.}\ \bibnamefont
  {Wang}}, \bibinfo {author} {\bibfnamefont {L.~S.}\ \bibnamefont {Wang}},
  \bibinfo {author} {\bibfnamefont {Y.~Y.}\ \bibnamefont {Guo}}, \bibinfo
  {author} {\bibfnamefont {Y.~C.}\ \bibnamefont {Wang}},\ and\ \bibinfo
  {author} {\bibfnamefont {A.~B.}\ \bibnamefont {Wang}},\ }\bibfield  {title}
  {\bibinfo {title} {Key space enhancement of optical chaos secure
  communication: chirped fbg feedback semiconductor laser},\ }\href
  {https://doi.org/10.1364/OE.27.003065} {\bibfield  {journal} {\bibinfo
  {journal} {Opt. Express}\ }\textbf {\bibinfo {volume} {27}},\ \bibinfo
  {pages} {3065} (\bibinfo {year} {2019})}\BibitemShut {NoStop}%
\bibitem [{\citenamefont {Chao}\ \emph {et~al.}(2020)\citenamefont {Chao},
  \citenamefont {Wang}, \citenamefont {Wang}, \citenamefont {Sun},
  \citenamefont {Han}, \citenamefont {Guo}, \citenamefont {Jia}, \citenamefont
  {Wang},\ and\ \citenamefont {Wang}}]{CHAO2020124702}%
  \BibitemOpen
  \bibfield  {author} {\bibinfo {author} {\bibfnamefont {M.}~\bibnamefont
  {Chao}}, \bibinfo {author} {\bibfnamefont {D.}~\bibnamefont {Wang}}, \bibinfo
  {author} {\bibfnamefont {L.}~\bibnamefont {Wang}}, \bibinfo {author}
  {\bibfnamefont {Y.}~\bibnamefont {Sun}}, \bibinfo {author} {\bibfnamefont
  {H.}~\bibnamefont {Han}}, \bibinfo {author} {\bibfnamefont {Y.}~\bibnamefont
  {Guo}}, \bibinfo {author} {\bibfnamefont {Z.}~\bibnamefont {Jia}}, \bibinfo
  {author} {\bibfnamefont {Y.}~\bibnamefont {Wang}},\ and\ \bibinfo {author}
  {\bibfnamefont {A.}~\bibnamefont {Wang}},\ }\bibfield  {title} {\bibinfo
  {title} {Permutation entropy analysis of chaotic semiconductor laser with
  chirped fbg feedback},\ }\href
  {https://doi.org/https://doi.org/10.1016/j.optcom.2019.124702} {\bibfield
  {journal} {\bibinfo  {journal} {Optics Communications}\ }\textbf {\bibinfo
  {volume} {456}},\ \bibinfo {pages} {124702} (\bibinfo {year}
  {2020})}\BibitemShut {NoStop}%
\bibitem [{\citenamefont {Ke}\ \emph {et~al.}(2017)\citenamefont {Ke},
  \citenamefont {Yi}, \citenamefont {Hou}, \citenamefont {Hu}, \citenamefont
  {Xia},\ and\ \citenamefont {Hu}}]{7882637}%
  \BibitemOpen
  \bibfield  {author} {\bibinfo {author} {\bibfnamefont {J.}~\bibnamefont
  {Ke}}, \bibinfo {author} {\bibfnamefont {L.}~\bibnamefont {Yi}}, \bibinfo
  {author} {\bibfnamefont {T.}~\bibnamefont {Hou}}, \bibinfo {author}
  {\bibfnamefont {Y.}~\bibnamefont {Hu}}, \bibinfo {author} {\bibfnamefont
  {G.}~\bibnamefont {Xia}},\ and\ \bibinfo {author} {\bibfnamefont
  {W.}~\bibnamefont {Hu}},\ }\bibfield  {title} {\bibinfo {title} {Time delay
  concealment in feedback chaotic systems with dispersion in loop},\ }\href
  {https://doi.org/10.1109/JPHOT.2017.2664341} {\bibfield  {journal} {\bibinfo
  {journal} {IEEE Photonics Journal}\ }\textbf {\bibinfo {volume} {9}},\
  \bibinfo {pages} {1} (\bibinfo {year} {2017})}\BibitemShut {NoStop}%
\bibitem [{\citenamefont {Zhang}\ \emph {et~al.}(2018)\citenamefont {Zhang},
  \citenamefont {Li}, \citenamefont {Wang}, \citenamefont {Zhang},
  \citenamefont {Ji},\ and\ \citenamefont {Wang}}]{Zhang:18}%
  \BibitemOpen
  \bibfield  {author} {\bibinfo {author} {\bibfnamefont {J.}~\bibnamefont
  {Zhang}}, \bibinfo {author} {\bibfnamefont {M.}~\bibnamefont {Li}}, \bibinfo
  {author} {\bibfnamefont {A.}~\bibnamefont {Wang}}, \bibinfo {author}
  {\bibfnamefont {M.}~\bibnamefont {Zhang}}, \bibinfo {author} {\bibfnamefont
  {Y.}~\bibnamefont {Ji}},\ and\ \bibinfo {author} {\bibfnamefont
  {Y.}~\bibnamefont {Wang}},\ }\bibfield  {title} {\bibinfo {title}
  {Time-delay-signature-suppressed broadband chaos generated by scattering
  feedback and optical injection},\ }\href
  {https://doi.org/10.1364/AO.57.006314} {\bibfield  {journal} {\bibinfo
  {journal} {Appl. Opt.}\ }\textbf {\bibinfo {volume} {57}},\ \bibinfo {pages}
  {6314} (\bibinfo {year} {2018})}\BibitemShut {NoStop}%
\bibitem [{\citenamefont {Th{\'e}venaz}\ \emph {et~al.}(2014)\citenamefont
  {Th{\'e}venaz}, \citenamefont {Chin}, \citenamefont {Sancho},\ and\
  \citenamefont {Sales}}]{10.1117/12.2059668}%
  \BibitemOpen
  \bibfield  {author} {\bibinfo {author} {\bibfnamefont {L.}~\bibnamefont
  {Th{\'e}venaz}}, \bibinfo {author} {\bibfnamefont {S.}~\bibnamefont {Chin}},
  \bibinfo {author} {\bibfnamefont {J.}~\bibnamefont {Sancho}},\ and\ \bibinfo
  {author} {\bibfnamefont {S.}~\bibnamefont {Sales}},\ }\bibfield  {title}
  {\bibinfo {title} {{Novel technique for distributed fibre sensing based on
  faint long gratings (FLOGs)}}\ }(\bibinfo  {publisher} {SPIE},\ \bibinfo
  {year} {2014})\ p.\ \bibinfo {pages} {91576W}\BibitemShut {NoStop}%
\bibitem [{\citenamefont {Yang}\ \emph {et~al.}(2022)\citenamefont {Yang},
  \citenamefont {Zhan}, \citenamefont {Cheng}, \citenamefont {Gan},
  \citenamefont {Fan},\ and\ \citenamefont {Tang}}]{Yang:22}%
  \BibitemOpen
  \bibfield  {author} {\bibinfo {author} {\bibfnamefont {M.}~\bibnamefont
  {Yang}}, \bibinfo {author} {\bibfnamefont {H.}~\bibnamefont {Zhan}}, \bibinfo
  {author} {\bibfnamefont {C.}~\bibnamefont {Cheng}}, \bibinfo {author}
  {\bibfnamefont {W.}~\bibnamefont {Gan}}, \bibinfo {author} {\bibfnamefont
  {D.}~\bibnamefont {Fan}},\ and\ \bibinfo {author} {\bibfnamefont
  {J.}~\bibnamefont {Tang}},\ }\bibfield  {title} {\bibinfo {title}
  {Large-capacity and long-distance distributed acoustic sensing based on an
  ultra-weak fiber bragg grating array with an optimized pulsed optical power
  arrangement},\ }\href {https://doi.org/10.1364/OE.455252} {\bibfield
  {journal} {\bibinfo  {journal} {Opt. Express}\ }\textbf {\bibinfo {volume}
  {30}},\ \bibinfo {pages} {16931} (\bibinfo {year} {2022})}\BibitemShut
  {NoStop}%
\bibitem [{\citenamefont {Yousefi}\ and\ \citenamefont
  {Lenstra}(1999)}]{766841}%
  \BibitemOpen
  \bibfield  {author} {\bibinfo {author} {\bibfnamefont {M.}~\bibnamefont
  {Yousefi}}\ and\ \bibinfo {author} {\bibfnamefont {D.}~\bibnamefont
  {Lenstra}},\ }\bibfield  {title} {\bibinfo {title} {Dynamical behavior of a
  semiconductor laser with filtered external optical feedback},\ }\href
  {https://doi.org/10.1109/3.766841} {\bibfield  {journal} {\bibinfo  {journal}
  {IEEE Journal of Quantum Electronics}\ }\textbf {\bibinfo {volume} {35}},\
  \bibinfo {pages} {970} (\bibinfo {year} {1999})}\BibitemShut {NoStop}%
\bibitem [{\citenamefont {Lang}\ and\ \citenamefont
  {Kobayashi}(1980)}]{1070479}%
  \BibitemOpen
  \bibfield  {author} {\bibinfo {author} {\bibfnamefont {R.}~\bibnamefont
  {Lang}}\ and\ \bibinfo {author} {\bibfnamefont {K.}~\bibnamefont
  {Kobayashi}},\ }\bibfield  {title} {\bibinfo {title} {External optical
  feedback effects on semiconductor injection laser properties},\ }\href
  {https://doi.org/10.1109/JQE.1980.1070479} {\bibfield  {journal} {\bibinfo
  {journal} {IEEE Journal of Quantum Electronics}\ }\textbf {\bibinfo {volume}
  {16}},\ \bibinfo {pages} {347} (\bibinfo {year} {1980})}\BibitemShut
  {NoStop}%
\bibitem [{\citenamefont {Donati}\ and\ \citenamefont {Horng}(2013)}]{6384650}%
  \BibitemOpen
  \bibfield  {author} {\bibinfo {author} {\bibfnamefont {S.}~\bibnamefont
  {Donati}}\ and\ \bibinfo {author} {\bibfnamefont {R.-H.}\ \bibnamefont
  {Horng}},\ }\bibfield  {title} {\bibinfo {title} {The diagram of feedback
  regimes revisited},\ }\href {https://doi.org/10.1109/JSTQE.2012.2234445}
  {\bibfield  {journal} {\bibinfo  {journal} {IEEE Journal of Selected Topics
  in Quantum Electronics}\ }\textbf {\bibinfo {volume} {19}},\ \bibinfo {pages}
  {1500309} (\bibinfo {year} {2013})}\BibitemShut {NoStop}%
\bibitem [{\citenamefont {Naumenko}\ \emph {et~al.}(2003)\citenamefont
  {Naumenko}, \citenamefont {Besnard}, \citenamefont {Loiko}, \citenamefont
  {Ughetto},\ and\ \citenamefont {Bertreux}}]{1233724}%
  \BibitemOpen
  \bibfield  {author} {\bibinfo {author} {\bibfnamefont {A.}~\bibnamefont
  {Naumenko}}, \bibinfo {author} {\bibfnamefont {P.}~\bibnamefont {Besnard}},
  \bibinfo {author} {\bibfnamefont {N.}~\bibnamefont {Loiko}}, \bibinfo
  {author} {\bibfnamefont {G.}~\bibnamefont {Ughetto}},\ and\ \bibinfo {author}
  {\bibfnamefont {J.}~\bibnamefont {Bertreux}},\ }\bibfield  {title} {\bibinfo
  {title} {Characteristics of a semiconductor laser coupled with a fiber bragg
  grating with arbitrary amount of feedback},\ }\href
  {https://doi.org/10.1109/JQE.2003.817669} {\bibfield  {journal} {\bibinfo
  {journal} {IEEE Journal of Quantum Electronics}\ }\textbf {\bibinfo {volume}
  {39}},\ \bibinfo {pages} {1216} (\bibinfo {year} {2003})}\BibitemShut
  {NoStop}%
\bibitem [{\citenamefont {Li}\ \emph {et~al.}(2014)\citenamefont {Li},
  \citenamefont {Wu}, \citenamefont {Zhong}, \citenamefont {Yang},
  \citenamefont {Mao},\ and\ \citenamefont {Xia}}]{Li:14}%
  \BibitemOpen
  \bibfield  {author} {\bibinfo {author} {\bibfnamefont {Y.}~\bibnamefont
  {Li}}, \bibinfo {author} {\bibfnamefont {Z.-M.}\ \bibnamefont {Wu}}, \bibinfo
  {author} {\bibfnamefont {Z.-Q.}\ \bibnamefont {Zhong}}, \bibinfo {author}
  {\bibfnamefont {X.-J.}\ \bibnamefont {Yang}}, \bibinfo {author}
  {\bibfnamefont {S.}~\bibnamefont {Mao}},\ and\ \bibinfo {author}
  {\bibfnamefont {G.-Q.}\ \bibnamefont {Xia}},\ }\bibfield  {title} {\bibinfo
  {title} {Time-delay signature of chaos in 1550 nm vcsels with
  variable-polarization fbg feedback},\ }\href
  {https://doi.org/10.1364/OE.22.019610} {\bibfield  {journal} {\bibinfo
  {journal} {Opt. Express}\ }\textbf {\bibinfo {volume} {22}},\ \bibinfo
  {pages} {19610} (\bibinfo {year} {2014})}\BibitemShut {NoStop}%
\bibitem [{\citenamefont {Erdogan}(1997)}]{618322}%
  \BibitemOpen
  \bibfield  {author} {\bibinfo {author} {\bibfnamefont {T.}~\bibnamefont
  {Erdogan}},\ }\bibfield  {title} {\bibinfo {title} {Fiber grating spectra},\
  }\href {https://doi.org/10.1109/50.618322} {\bibfield  {journal} {\bibinfo
  {journal} {Journal of Lightwave Technology}\ }\textbf {\bibinfo {volume}
  {15}},\ \bibinfo {pages} {1277} (\bibinfo {year} {1997})}\BibitemShut
  {NoStop}%
\bibitem [{\citenamefont
  {Schuster}(2005)}]{doi:https://doi.org/10.1002/3527604804.ch6}%
  \BibitemOpen
  \bibfield  {author} {\bibinfo {author} {\bibfnamefont {H.~G.}\ \bibnamefont
  {Schuster}},\ }\bibinfo {title} {Strange attractors in dissipative dynamical
  systems},\ in\ \href {https://doi.org/https://doi.org/10.1002/3527604804.ch6}
  {\emph {\bibinfo {booktitle} {Deterministic Chaos}}}\ (\bibinfo  {publisher}
  {John Wiley \& Sons, Ltd},\ \bibinfo {year} {2005})\ Chap.~\bibinfo {chapter}
  {6}, pp.\ \bibinfo {pages} {89--125}\BibitemShut {NoStop}%
\bibitem [{\citenamefont {Eckhardt}\ and\ \citenamefont
  {Yao}(1993)}]{ECKHARDT1993100}%
  \BibitemOpen
  \bibfield  {author} {\bibinfo {author} {\bibfnamefont {B.}~\bibnamefont
  {Eckhardt}}\ and\ \bibinfo {author} {\bibfnamefont {D.}~\bibnamefont {Yao}},\
  }\bibfield  {title} {\bibinfo {title} {Local lyapunov exponents in chaotic
  systems},\ }\href
  {https://doi.org/https://doi.org/10.1016/0167-2789(93)90007-N} {\bibfield
  {journal} {\bibinfo  {journal} {Physica D: Nonlinear Phenomena}\ }\textbf
  {\bibinfo {volume} {65}},\ \bibinfo {pages} {100} (\bibinfo {year}
  {1993})}\BibitemShut {NoStop}%
\bibitem [{\citenamefont {Fischer}\ \emph {et~al.}(2004)\citenamefont
  {Fischer}, \citenamefont {Yousefi}, \citenamefont {Lenstra}, \citenamefont
  {Carter},\ and\ \citenamefont {Vemuri}}]{1366365}%
  \BibitemOpen
  \bibfield  {author} {\bibinfo {author} {\bibfnamefont {A.}~\bibnamefont
  {Fischer}}, \bibinfo {author} {\bibfnamefont {M.}~\bibnamefont {Yousefi}},
  \bibinfo {author} {\bibfnamefont {D.}~\bibnamefont {Lenstra}}, \bibinfo
  {author} {\bibfnamefont {M.}~\bibnamefont {Carter}},\ and\ \bibinfo {author}
  {\bibfnamefont {G.}~\bibnamefont {Vemuri}},\ }\bibfield  {title} {\bibinfo
  {title} {Experimental and theoretical study of semiconductor laser dynamics
  due to filtered optical feedback},\ }\href
  {https://doi.org/10.1109/JSTQE.2004.835997} {\bibfield  {journal} {\bibinfo
  {journal} {IEEE Journal of Selected Topics in Quantum Electronics}\ }\textbf
  {\bibinfo {volume} {10}},\ \bibinfo {pages} {944} (\bibinfo {year}
  {2004})}\BibitemShut {NoStop}%
\bibitem [{\citenamefont {Erzgr\"aber}\ \emph {et~al.}(2007)\citenamefont
  {Erzgr\"aber}, \citenamefont {Lenstra}, \citenamefont {Krauskopf},
  \citenamefont {Fischer},\ and\ \citenamefont {Vemuri}}]{PhysRevE.76.026212}%
  \BibitemOpen
  \bibfield  {author} {\bibinfo {author} {\bibfnamefont {H.}~\bibnamefont
  {Erzgr\"aber}}, \bibinfo {author} {\bibfnamefont {D.}~\bibnamefont
  {Lenstra}}, \bibinfo {author} {\bibfnamefont {B.}~\bibnamefont {Krauskopf}},
  \bibinfo {author} {\bibfnamefont {A.~P.~A.}\ \bibnamefont {Fischer}},\ and\
  \bibinfo {author} {\bibfnamefont {G.}~\bibnamefont {Vemuri}},\ }\bibfield
  {title} {\bibinfo {title} {Feedback phase sensitivity of a semiconductor
  laser subject to filtered optical feedback: Experiment and theory},\ }\href
  {https://doi.org/10.1103/PhysRevE.76.026212} {\bibfield  {journal} {\bibinfo
  {journal} {Phys. Rev. E}\ }\textbf {\bibinfo {volume} {76}},\ \bibinfo
  {pages} {026212} (\bibinfo {year} {2007})}\BibitemShut {NoStop}%
\bibitem [{\citenamefont {Sk\"{e}nderas}\ \emph {et~al.}(2022)\citenamefont
  {Sk\"{e}nderas}, \citenamefont {Jolly}, \citenamefont {Gupta}, \citenamefont
  {Geernaert},\ and\ \citenamefont {Virte}}]{Skenderas:22}%
  \BibitemOpen
  \bibfield  {author} {\bibinfo {author} {\bibfnamefont {M.}~\bibnamefont
  {Sk\"{e}nderas}}, \bibinfo {author} {\bibfnamefont {S.~W.}\ \bibnamefont
  {Jolly}}, \bibinfo {author} {\bibfnamefont {N.}~\bibnamefont {Gupta}},
  \bibinfo {author} {\bibfnamefont {T.}~\bibnamefont {Geernaert}},\ and\
  \bibinfo {author} {\bibfnamefont {M.}~\bibnamefont {Virte}},\ }\bibfield
  {title} {\bibinfo {title} {Impact of {FBG} feedback phase on laser
  dynamics},\ }\href {https://doi.org/10.1364/OL.451598} {\bibfield  {journal}
  {\bibinfo  {journal} {Opt. Lett.}\ }\textbf {\bibinfo {volume} {47}},\
  \bibinfo {pages} {1602} (\bibinfo {year} {2022})}\BibitemShut {NoStop}%
\bibitem [{\citenamefont {Uchida}(2012)}]{uchida2012optical}%
  \BibitemOpen
  \bibfield  {author} {\bibinfo {author} {\bibfnamefont {A.}~\bibnamefont
  {Uchida}},\ }\href@noop {} {\emph {\bibinfo {title} {Optical communication
  with chaotic lasers: applications of nonlinear dynamics and
  synchronization}}}\ (\bibinfo  {publisher} {John Wiley \& Sons},\ \bibinfo
  {year} {2012})\BibitemShut {NoStop}%
\bibitem [{\citenamefont {M\o{}rk}\ \emph {et~al.}(1990)\citenamefont
  {M\o{}rk}, \citenamefont {Mark},\ and\ \citenamefont
  {Tromborg}}]{PhysRevLett.65.1999}%
  \BibitemOpen
  \bibfield  {author} {\bibinfo {author} {\bibfnamefont {J.}~\bibnamefont
  {M\o{}rk}}, \bibinfo {author} {\bibfnamefont {J.}~\bibnamefont {Mark}},\ and\
  \bibinfo {author} {\bibfnamefont {B.}~\bibnamefont {Tromborg}},\ }\bibfield
  {title} {\bibinfo {title} {Route to chaos and competition between relaxation
  oscillations for a semiconductor laser with optical feedback},\ }\href
  {https://doi.org/10.1103/PhysRevLett.65.1999} {\bibfield  {journal} {\bibinfo
   {journal} {Phys. Rev. Lett.}\ }\textbf {\bibinfo {volume} {65}},\ \bibinfo
  {pages} {1999} (\bibinfo {year} {1990})}\BibitemShut {NoStop}%
\bibitem [{\citenamefont {Cohen}\ \emph {et~al.}(1988)\citenamefont {Cohen},
  \citenamefont {Drenten},\ and\ \citenamefont {Verbeeck}}]{8533}%
  \BibitemOpen
  \bibfield  {author} {\bibinfo {author} {\bibfnamefont {J.}~\bibnamefont
  {Cohen}}, \bibinfo {author} {\bibfnamefont {R.}~\bibnamefont {Drenten}},\
  and\ \bibinfo {author} {\bibfnamefont {B.}~\bibnamefont {Verbeeck}},\
  }\bibfield  {title} {\bibinfo {title} {The effect of optical feedback on the
  relaxation oscillation in semiconductor lasers},\ }\href
  {https://doi.org/10.1109/3.8533} {\bibfield  {journal} {\bibinfo  {journal}
  {IEEE Journal of Quantum Electronics}\ }\textbf {\bibinfo {volume} {24}},\
  \bibinfo {pages} {1989} (\bibinfo {year} {1988})}\BibitemShut {NoStop}%
\bibitem [{\citenamefont {Mork}\ \emph {et~al.}(1992)\citenamefont {Mork},
  \citenamefont {Tromborg},\ and\ \citenamefont {Mark}}]{119502}%
  \BibitemOpen
  \bibfield  {author} {\bibinfo {author} {\bibfnamefont {J.}~\bibnamefont
  {Mork}}, \bibinfo {author} {\bibfnamefont {B.}~\bibnamefont {Tromborg}},\
  and\ \bibinfo {author} {\bibfnamefont {J.}~\bibnamefont {Mark}},\ }\bibfield
  {title} {\bibinfo {title} {Chaos in semiconductor lasers with optical
  feedback: theory and experiment},\ }\href {https://doi.org/10.1109/3.119502}
  {\bibfield  {journal} {\bibinfo  {journal} {IEEE Journal of Quantum
  Electronics}\ }\textbf {\bibinfo {volume} {28}},\ \bibinfo {pages} {93}
  (\bibinfo {year} {1992})}\BibitemShut {NoStop}%
\bibitem [{\citenamefont {Lenstra}(2013)}]{6459532}%
  \BibitemOpen
  \bibfield  {author} {\bibinfo {author} {\bibfnamefont {D.}~\bibnamefont
  {Lenstra}},\ }\bibfield  {title} {\bibinfo {title} {Relaxation oscillation
  dynamics in semiconductor diode lasers with optical feedback},\ }\href
  {https://doi.org/10.1109/LPT.2013.2246562} {\bibfield  {journal} {\bibinfo
  {journal} {IEEE Photonics Technology Letters}\ }\textbf {\bibinfo {volume}
  {25}},\ \bibinfo {pages} {591} (\bibinfo {year} {2013})}\BibitemShut
  {NoStop}%
\bibitem [{\citenamefont {Liu}\ \emph {et~al.}(2021)\citenamefont {Liu},
  \citenamefont {Ruan}, \citenamefont {Yu}, \citenamefont {Wang},\ and\
  \citenamefont {An}}]{Liu:21}%
  \BibitemOpen
  \bibfield  {author} {\bibinfo {author} {\bibfnamefont {B.}~\bibnamefont
  {Liu}}, \bibinfo {author} {\bibfnamefont {Y.}~\bibnamefont {Ruan}}, \bibinfo
  {author} {\bibfnamefont {Y.}~\bibnamefont {Yu}}, \bibinfo {author}
  {\bibfnamefont {B.}~\bibnamefont {Wang}},\ and\ \bibinfo {author}
  {\bibfnamefont {L.}~\bibnamefont {An}},\ }\bibfield  {title} {\bibinfo
  {title} {Influence of feedback optical phase on the relaxation oscillation
  frequency of a semiconductor laser and its application},\ }\href
  {https://doi.org/10.1364/OE.414849} {\bibfield  {journal} {\bibinfo
  {journal} {Opt. Express}\ }\textbf {\bibinfo {volume} {29}},\ \bibinfo
  {pages} {3163} (\bibinfo {year} {2021})}\BibitemShut {NoStop}%
\bibitem [{\citenamefont {Chan}(2010)}]{5419246}%
  \BibitemOpen
  \bibfield  {author} {\bibinfo {author} {\bibfnamefont {S.-C.}\ \bibnamefont
  {Chan}},\ }\bibfield  {title} {\bibinfo {title} {Analysis of an optically
  injected semiconductor laser for microwave generation},\ }\href
  {https://doi.org/10.1109/JQE.2009.2028900} {\bibfield  {journal} {\bibinfo
  {journal} {IEEE Journal of Quantum Electronics}\ }\textbf {\bibinfo {volume}
  {46}},\ \bibinfo {pages} {421} (\bibinfo {year} {2010})}\BibitemShut
  {NoStop}%
\bibitem [{\citenamefont {Grillot}\ \emph {et~al.}(2008)\citenamefont
  {Grillot}, \citenamefont {Dagens}, \citenamefont {Provost}, \citenamefont
  {Su},\ and\ \citenamefont {Lester}}]{4633714}%
  \BibitemOpen
  \bibfield  {author} {\bibinfo {author} {\bibfnamefont {F.}~\bibnamefont
  {Grillot}}, \bibinfo {author} {\bibfnamefont {B.}~\bibnamefont {Dagens}},
  \bibinfo {author} {\bibfnamefont {J.-G.}\ \bibnamefont {Provost}}, \bibinfo
  {author} {\bibfnamefont {H.}~\bibnamefont {Su}},\ and\ \bibinfo {author}
  {\bibfnamefont {L.~F.}\ \bibnamefont {Lester}},\ }\bibfield  {title}
  {\bibinfo {title} {Gain compression and above-threshold linewidth enhancement
  factor in 1.3-$\mu\hbox{m}$ inas–gaas quantum-dot lasers},\ }\href
  {https://doi.org/10.1109/JQE.2008.2003106} {\bibfield  {journal} {\bibinfo
  {journal} {IEEE Journal of Quantum Electronics}\ }\textbf {\bibinfo {volume}
  {44}},\ \bibinfo {pages} {946} (\bibinfo {year} {2008})}\BibitemShut
  {NoStop}%
\end{thebibliography}%

\end{document}